\documentclass[a4paper,10pt]{article}

\usepackage[dvips]{graphicx}
\usepackage{epsfig}
\usepackage{graphicx}
\usepackage{subcaption}
\usepackage{amssymb}
\usepackage{amsmath}
\usepackage{textcomp}
\usepackage{cite}
\usepackage{subcaption}
\def\be{\begin{equation}}

\def\ee{\end{equation}}
\begin{document}
\title{Wormhole inducing inflation with Einstein Gauss Bonnet dilaton interaction }
\author{ Gargi Biswas$^{}$\footnote{E-mail: biswasgb@gmail.com}, M K Dutta$^{}$\footnote{E-mail:mkdbrsnc@gmail.com } and B Modak$^{}$\footnote{E-mail: bijanmodak@yahoo.co.in}~}

\maketitle
\noindent
\begin{center}
\noindent
$^{\ast,\ddag}$ Department of Physics, University of Kalyani, Kalyani-741235,India.\\
$^{\dag}$ Department of Physics, BRSN College, Kolkata- 740120, India.\\
\end{center}

\begin{abstract}
A few Euclidean wormhole configurations are presented using both the analytic and numerical solutions of the field equations in 4-dimensional Robertson Walker Euclidean background with Einstein Gauss-Bonnet dilaton interaction. In one analytic solution we present transition from a wormhole to an exponential expansion with Lorentzian time $t$ using $\tau=it$ after passing through an era of oscillating  Euclidean wormhole. The numerical solutions of the scale factor $a(\tau)$ show multiple local maxima and minima about a global minimum for inverse power law potentials, while for exponential potential the wormholes have a single minimum. An inflationary cosmic scenario away from the throat of the wormhole can be obtained from the Hubble parameter and deceleration parameter obtained by curve fit of $a(\tau)$ of the numerical solution invoking analytic continuation by $\tau=it$. The potential is also observed to decay sharply.
\end{abstract}
\section{Introduction:}
Inflationary cosmology \cite{gu:prd,li:pl,stei:prl,a:plb} resolves most of the problems of standard cosmology, except cosmic singularity and vanishing of cosmological constant \cite{pd:book}. To avoid cosmic singularity in the early universe several gravitational theories, like modified theories of gravity \cite{stell:prd, capp:prd, harko:prd, lobo:fR}, string theory \cite{sloan:string, met:string}, Kaluza-Klein gravity\cite{Dzhu:prd,ks:fR}, etc appear as natural generalization of Einstein's gravity. The Einstein Gauss Bonnet dilatonic (referred as DEGB) coupled interaction arises logically as the leading order in the low energy effective Heterotic string theory \cite{sloan:string, met:string}. As a consequence, this interaction works satisfactorily to explain inflation \cite{Neu:jcap} as well as late-time acceleration \cite{sami:jcap} and is used also to study possible resolution of initial singularity  \cite{Kanti:prd, Anto:np}. The singularity problem is still illusive, whose resolution has not yet become visible. The Euclidean wormhole \cite{eu:wh1,eu:wh2} is fascinating to alleviate initial cosmic singularity problem as well as to render a viable mechanism of vanishing of cosmological constant, since it represents tunnelling of the universe through the classical singularity.
\par
In the early nineties Coleman \cite{co:nu}, Baum \cite{baum:pl} and Hawking \cite{haw:pl} proposed a mechanism of vanishing of cosmological constant in quantum cosmology introducing an idea of wormhole at the Planck era. Interestingly, Giddings and Strominger \cite{gs:np} first proposed the idea of wormhole from a solution of Euclidean field equations in the Einstein's gravity with axionic field. In continuation, numerous solutions \cite{morris1,morris2,hall:cqg,cou:cqg,lobo:bd,nandi:bd} have been explored in different fields, as well as consequences of them were also extensively followed up \cite{co:nu}, \cite{haw:prd,gid:npl, fisu:plb, unr:prd,hawk:npb, hawk:page,kle:npb}. The null energy condition (NEC) is violated in the wormhole solutions \cite{visser:nec}, however it can be minimised in the alternative theories of gravity \cite{capp:prd,harko:prd,lobo:fR,sloan:string,met:string,Dzhu:prd,ks:fR} replacing axionic field by the additional geometric contribution.

\par
Cosmic evolution from the Planck era to the inflationary era is not smooth as the pre-inflationary era is governed by quantum gravity in curved spacetime \cite{qg:wh,qft:bd}, while inflationary era can be described by the field theory. Further, it is thought that the spacetime is Euclidean in the early era, wherein the Euclidean time $\tau$ plays the role of time, subsequently the spacetime is Lorentzian and the time $t$ is the cosmic time. The Euclidean domain is classically forbidden in general and one cannot probe it with $t$, thus using Wick rotation $\tau=it$ one can analytically continue a function from one region to another or vice versa. Intuitively, the wormhole, if it exists at all in the Planck era, then  subsequent evolution  naturally demands an era of inflation. Lorentzian wormholes which represent non-trivial geometry in interstellar space have been considered  in the DEGB theory \cite{kanti:GB,Mehdi:GB,bhawal:GB}, however viability of an inflation from a wormhole configuration has not been confirmed. In recent works, \cite{gb:slowroll,R:Comm,GBbm:comm} transition from wormhole configuration to inflationary era is evident by using $\tau=it$. A transition from wormhole solution to inflation is confirmed in \cite{gb:slowroll} with  numerical solutions in the DEGB theory with a few power law potentials. So we extend our earlier work of wormhole with exponential and inverse power law potentials \cite{ratra:inverse} in the DEGB theory. In the extended form of our earlier work, wormhole of multiple maxima and minima  about the global minimum  \cite{gb:slowroll} emerges in the Euclidean space with large class of inverse power law potential.
\par
We consider both the analytic and numerical solutions of wormhole in the DEGB theory in 4-dimensional Robertson Walker Euclidean background. Analytic solutions are obtained with a simplifying assumption on the dynamical coupling $\Lambda(\phi)$ and with some constraints on the potential $V(\phi)$. Analytic solutions show wormhole configuration in the early era of $\tau$. In one case we get an inflationary scenario asymptotically like \cite{R2:bm} after crossing an era of oscillating Euclidean universe beginning from a Euclidean wormhole solution.

\par
The numerical wormhole solutions satisfying all the field equations like our work \cite{gb:slowroll} are considered with a few potentials favourable for inflation \cite{ratra:inverse,ak:cqg,5:GB}. These solutions are presented with a plot of $a(\tau)$ as a function of  $\tau$, which show that two asymptotic domains are connected by a tube of finite radius at the minimum of $a(\tau)$. The solution shows multiple maxima and minima unlike \cite{gb:slowroll} about  global minimum for inverse power law potentials. By and large, $a(\tau)$  increases with $ \vert \tau\vert$ superimposed with tiny oscillation for inverse potentials. Now to explore the cosmic scenario of the numerical solutions we consider curve fit of $a(\tau)$ as polynomial of $\tau$ and consequently $a(t)$ is obtained by using $\tau=it$. Thereby $a(t)$ is used to find the observable, viz. the Hubble parameter $H(t)$ \big( or $H(z)$\big) and the deceleration parameter $q(t)$ \big( or $q(z)$\big). These are then used to investigate dynamical scenario of the wormhole, hence cosmic evolution from a plot of $H(t)$ and $q(t)$ with $t$. Further, using the absolute value of the scale factor $a(t)$ the plot of $H(z)$ and $q(z)$ with red shift parameter $z$ are considered.
\par
Evolution of $H(t)$ and $q(t)$ show an initial collapsing phase till  $t=t_i$ before encountering some unusual evolution around the throat of the wormhole followed by final expansion initiating at $t=t_f$. The domain $(t_f-t_i)$ shows an unusual evolution, which is almost similar to the classical forbidden domain in the analytic solution in \cite{R:Comm}. Asymptotically $H(t)$ approaches to a constant value at $t>>t_f$, wherein $q(t) \rightarrow -1$. Asymptotic values of $H(z)$ and $q(z)$ at the extreme values of $z$ also confirm final inflationary era. Thus a phase transition from a Euclidean space to Lorentzian space with the Wick rotation $\tau=it$ yields an inflationary era beginning from a wormhole \cite{inf:1,inf:2,inf:3}. The evolution appears to be consistent with the classical scenario away from the throat. The $V(\phi)$ decreases substantially to a very small value far away from the throat. The NEC is satisfied in all analytic and numerical solutions in the neighbourhood of the throat, while in the numerical solution it is also satisfied far away from the throat.
\par
We present the field equations in section 2. In section 3 analytic solutions are considered. Numerical solutions of the field equations are presented in section 4. We interpret the numerical solutions in section 5. Section 6 contains a brief discussion and finally an Appendix is given in section 7.

\section{Action with Einstein Gauss-Bonnet dilaton interaction and the field equations:}
We consider Gauss Bonnet dilaton interaction in the  Einstein Hilbert term as
\begin{equation}\label{g2.1}
S=\int {d^4}x \sqrt{g}\Big[\frac{R}{2K^2} - \frac{\gamma}{2} \phi_{;\mu}\phi^{;\mu} - V(\phi) - \frac{\Lambda(\phi)}{8}{\cal{G}}\Big] + S_{m},
\end{equation}
where  ${\cal{G}} = R_{\mu\nu\alpha\beta}R^{\mu\nu\alpha\beta} - 4R_{\mu\nu}R^{\mu\nu} + R^2$ is the Gauss-Bonnet curvature, R is the Ricci scalar, $\Lambda(\phi)$ is the dynamical coupling of the GB term with the dilaton field $\phi$ \cite{5:GB,gr:qc},  $V(\phi)$ is the potential, $K$ is the inverse of Planck mass, $\gamma$ is a constant and $S_{m}$ is the surface term.  In Robertson Walker Euclidean background
\[
ds^2 = d{\tau}^2 + a^2(\tau)\Big[\frac{dr^2}{1-\kappa r^2}+r^2 ( d{\theta}^2 + \sin^2\theta d\phi^2)\Big],
\]
the field equations are
\begin{equation}\label{g2.2}
\frac{3}{K^2}\Big( \frac{a'^{2}}{a^2} -\frac{\kappa}{a^2}\Big)=\frac{\gamma}{2}{\phi'}^2 - V(\phi) - 3{\Lambda}'~ \frac{a'}{a} \Big(\frac{a'^{2}}{a^2} -\frac{\kappa}{a^2}\Big),
\end{equation}

\begin{equation}\label{g2.3}
-\frac{1}{K^2}\Big(2 \frac{{a''}}{a}+\frac{{a'}^2}{a^2}-\frac{\kappa}{a^2}\Big) = \frac{\gamma}{2}{\phi'}^2 + V(\phi)+\Lambda''\Big(\frac{{a'}^2}{a^2}-\frac{\kappa}{a^2}\Big)+2\Lambda'\frac{{a''}{a'}}{a^2}
\end{equation}
and
\begin{equation}\label{g2.4}
\gamma \Big(\phi''+3\frac{a'}{a}\phi'\Big) = V_{,\phi} + 3\Lambda_{,\phi}\frac{a''}{a}\Big(\frac{a'^{2}}{a^2}-\frac{\kappa}{a^2}\Big)=\frac{\partial V_{eff}}{\partial \phi},
\end{equation}
where $a(\tau)$ is the scale factor and $\frac{\partial V_{eff}}{\partial \phi}$ is the effective potential gradient having contribution from the potential $V(\phi)$ and GB dynamical coupling $\Lambda(\phi)$. Here, $\kappa$ is the 3-space curvature parameter $(\kappa=0,\pm1)$. A prime denotes derivative with respect to Euclidean time $\tau$, whereas a comma $(,)$ denotes partial derivative. Evolution of the universe is determined by \eqref{g2.2}-\eqref{g2.4} for dynamical coupling $\Lambda(\phi)$ and potential $V(\phi)$.
\par
A wormhole should satisfy following conditions to be a solution of the field equations. A wormhole has two asymptotic regions connected by a tube with a non-vanishing minimum of the scale factor $a(\tau)$ at the throat at $\tau=\tau_0$, where $a'(\tau_0)=0$, $a''(\tau_0) > 0$, otherwise $a(\tau)$ is finite at other $\tau$. Then the equation \eqref{g2.2} at $\tau=\tau_0$ yields
\begin{equation}\label{g2.7}
\frac{3}{K^2}\frac{\kappa}{a_0^2}=-\frac{\gamma}{2}\phi_0'^2 +V_0,
\end{equation}
where a subscript ``$0$'' on a variable denotes the value of the variable at $\tau_0$.
So from \eqref{g2.7} the potential energy $V_0$  at the throat is very large positive as $ a_0 $ is small therein with $\kappa =1$. Further from \eqref{g2.3} and \eqref{g2.7}, $a''(\tau)$ at the extrema is
\begin{equation}\label{g2.8}
a''_0=\frac{K^2}{3}(-\gamma\phi'^2_0-V_0) a_0 +\frac{\kappa K^2}{2}\frac{\Lambda_0''}{a_0}.
\end{equation}
The Gauss-Bonnet contribution is assumed to be a small correction to the gravity, however the term $\frac{\Lambda_0''}{a_0}$ in \eqref{g2.8} may be large and it may take all possible values; so the scale factor may have multiple maxima and minima. Thus to ensure lower bound of $a(\tau)$ one should satisfy
$\phi_0'^2 > V_0$ for $\gamma=-1$ and $\kappa =1$ assuming $ \frac{\Lambda_0''}{a_0}>0$ and $V_0>0$. On the other hand, we may have local maxima or minima of $a(\tau)$ depending on the magnitude and the sign of $ \frac{\Lambda_0''}{a_0}$ and $V_0$ at the corresponding local extrema. The field equations may allow solution for $\kappa=0$ with $\gamma=-1$.
\\ Analytic solution is quite non-trivial in general with specific $V(\phi)$ and $\Lambda(\phi)$ for $\kappa \neq 0$. So we present a few analytic solutions with simplifying assumption for $\kappa = 0$ in the next section.

\section{Analytic solution of the field equations for wormhole configuration:}
Analytic solution is quite non trivial, so we consider assumption to get a glimpse of evolution in the very early universe. The solution of the field equations requires knowledge of $\Lambda(\phi)$ and $V(\phi)$. We simplify the field equations with a restriction on the coupling function $\Lambda(\phi)$ as

\begin{equation}\label{g2.8a}
 \Lambda'\frac{a'}{a}=\frac{2m}{K^2},
\end{equation}
where $m$ is a constant. In some earlier works \cite{la:mkd,la:rnb} the condition \eqref{g2.8a} has been used to study late time acceleration in the DEGB theory.
Now using \eqref{g2.8a} in combination of \eqref{g2.2} and \eqref{g2.3} for $\kappa=0$ we get
\begin{equation}\label{g2.9}
-2(m+1)a''a +(4m+2)a'^{2}=\gamma \phi'^2 a^2 K^2 ~~~~\hbox{and}
\end{equation}
\begin{equation}\label{g2.10}
(m+1)a''a + (4m+2)a'^{2}= -V(\phi) a^2 K^2.
\end{equation}
Now from \eqref{g2.9} and \eqref{g2.10} the condition of non-vanishing minimum radius of the wormhole, i.e. $a'_0=0$ and $a''_0 >0 $ at the throat at some time $\tau_0$ can be satisfied assuming
$\gamma=-1$ for $m+1>0$ with real scalar field, but with $V(\phi_0)<0$ at the throat, which yields the conditions $\frac{a''_0}{a_0}=\frac{K^2 \phi'^2_0}{2(m+1)}$ and $\frac{a''_0}{a_0}=-\frac{K^2 V(\phi_0)}{m+1}$.
The form of \eqref{g2.9} and \eqref{g2.10} are simple for solution with a knowledge of $V(\phi)$. Now we consider a few solutions with choice of $V(\phi)$.

\subsection{Solution with vanishing effective potential gradient:}
The contribution of Gauss Bonnet coupling term in the field equations gives rise to an additional potential apart from $V(\phi)$. The term in the right side of \eqref{g2.4} is an effective potential gradient $ \frac{\partial V_{eff}}{\partial \phi} $ and the dilaton field evolved under influence of this potential gradient. Instead of considering form of $V(\phi)$ we simplify the field equations assuming that the dilaton field is free from the influence of effective potential gradient, then \eqref{g2.4} for $\kappa=0$ leads to

\begin{equation}\label{g2.11}
\phi'a^3= c_0,~ \hbox{ and }~ V'= - 3 \Lambda' \frac{a''a'^2}{a^3•},
\ee
from which we can determine the potential, where $c_0$ is a constant. Now using above $\phi'$ with $\Lambda'\frac{a'}{a}=\frac{2m}{K^2}$ and $\gamma=-1$ in \eqref{g2.9} we have
\begin{equation}\label{g2.12}
(m+1)\frac{a''}{a} - (2m+1) \frac{a'^2}{a^2}= \frac{K^2 \phi'^2}{2}=\frac{K^2 c_0^2}{2a^6}.
\end{equation}
The first integral of \eqref{g2.12} gives
\begin{equation}\label{g2.13}
a^{4}a'^2 = -\frac{K^2 c_0^2}{8m+6}  + r_0^2 a ^{\frac{8m+6}{m+1}},
\end{equation}
where $r_0$ is a constant.The equation \eqref{g2.13} allows  different solutions for each $m$. Now for existence of wormhole $a'=0$ at the throat at some $\tau_0$, wherein $a_0''>0$. So at the throat of radius $a_0$

\begin{equation}\label{g2.13a}
a_0^{\frac{8m+6}{m+1}}=\frac{K^2 c_0^2}{(8m+6)r_0^2}~  \hbox{and } ~a_0'' =\frac{K^2 c_0^2}{2(m+1)a_0^5},
\end{equation}
so for wormhole solution $(4m+3) > 0$ (i.e. $m> -\frac{3}{4}$).
The null energy condition (NEC) from  \eqref{g2.2} and \eqref{g2.3} using \eqref{g2.13}can be expressed as
\begin{equation}\label{g2.13b}
\rho_{_{eff}}+p_{_{eff}}= \frac{2mr_0^2}{(m+1)K^2}a^{\frac{2m}{m+1}}+ \frac{6c_0^2}{(8m+6) a^6}= \frac{c_0^2}{(4m+3)a^6 }\Big[3 + \frac{m}{m+1} \Big(\frac{a}{a_0}\Big)^{\frac{8m+6}{m+1}} \Big],
\end{equation}
thus NEC shows that the energy condition may be satisfied near the throat with suitable $m$. To get an explicit evolution with $\tau$ we choose the value of $m$.
\subsubsection{Evolution of the universe for $m=-\frac{2}{3}$:}

A simple choice of $m$  yields solution of \eqref{g2.12} as
\begin{equation}\label{g2.14}
\frac{a}{2} \sqrt{a^2-a_0^2}+\frac{a_0^2}{2} \log\Big[a +\sqrt{a^2-a_0^2} \Big]= \pm r_0 \tau + r_1,
\end{equation}
where, $m=-\frac{2}{3}$ and $r_1$ is an integration constant. Further $r_1$ can be determined using the idea that the radius of throat $a(\tau)=a_0$ occurs at $\tau=0$, so $r_1=\frac{a_0^2}{2} \log[a_0] $, where $a_0 = \sqrt{\frac{3}{2}}\frac{Kc_0}{r_0}$. The solution \eqref{g2.14} gives rise to evolution for all $\tau$. Near the throat (i.e. $a \sim a_0$) second term in left side of \eqref{g2.14} is dominating, so
\begin{equation}\label{g2.15}
a(\tau)= a_0 \cosh(\omega\tau),
\end{equation}
while the first term in
\eqref{g2.14} is dominating in the regime $a>>a_0$. So to study evolution in $a>>a_0$, we consider solution in the cosmic time $t$ using $\tau=it$ in \eqref{g2.11}, which gives $\dot{\phi}a^3=ic_0=c_1$, where  $c_1$ is real for real $\phi$ in the Lorentz signature. Further, \eqref{g2.13} reduces to
\begin{equation}\label{g2.15a}
a^4\dot{a}^2= \frac{3 K^2 c_0^2}{2}-r_0^2 a^2,
\end{equation}
since the radius $a_0$ at the throat satisfies  \eqref{g2.13a} as $a_0^2= \frac{3 K^2 c_0^2}{2r_0^2} = -\frac{3 K^2 c_1^2}{2r_0^2}$, thus $r_0$ must be in the form $r_0=ir_1$, where $r_1$ is real, then $a_0^2= \frac{3 K^2 c_1^2}{2r_1^2} $ gives a real value. So  \eqref{g2.15a} reduces to
\begin{equation}\label{g2.15aa}
a^4\dot{a}^2= r_1^2(a^2-a_0^2).
\end{equation}. Hence in the domain $a>>a_0$, \eqref{g2.15aa} gives $a^2\dot{a}^2 \sim r_1^2$, so $a(t)\sim \sqrt{t}$. Thus an initial wormhole configuration evolves to a radiation dominated era asymptotically.
The potential $V(\phi)$ in this case leads to
\begin{equation}\label{g2.16}
V(a)= \frac{c_0^2}{a^6•}\Big[\frac{3}{2} \frac{a^2}{a_0^2•}-2 \Big],
\end{equation}
which is positive as long as  $a(\tau)> \frac{2a_0}{\sqrt{3}}$.
Further \eqref{g2.13b} now reduces to
\begin{equation}
\label{g2.13c}
\rho_{_{eff}}+p_{_{eff}}= \frac{4r_0^2}{K^2a^6}\Big(\frac{3a_0^2}{2}-a^2\Big) = \frac{9c_0^2}{a^6}\Big(1-\frac{2a^2}{3a_0^2}\Big) .
\end{equation}
Thus NEC is satisfied near the throat as long as $a(\tau) < \sqrt{\frac{3}{2}}a_0 $.

\subsection{Solution with a choice of potential of the dilaton field:}
The solution of the field equation with standard form of potential is non-trivial, so we assume potential $V$ as a function of the scale factor $a(\tau)$ \cite{carl:m} in the form
\begin{equation}\label{g2.17}
V(\phi(\tau))= \frac{v_0}{a^{\alpha}} -u_0.
\end{equation}
The choice of potential seems artificial, however it is possible  in the spatial homogeneous section, where $u_0$, $v_0$ and $\alpha $ are constants. Now introducing \eqref{g2.17} in \eqref{g2.10} and with an integration we get
\begin{equation}\label{g2.18}
a'^2 = - \frac{2v_0 K^2 ~a^{2-\alpha}}{[10m +6 -\alpha(m+1)]} + \frac{2u_0K^2 a^2 }{10m +6} + c_2 a^{-4\frac{(2m+1)}{(m+1)}},
\end{equation}
where $c_2$ is an integration constant. We have  class of solutions depending on $m$, $v_0$, $u_0$, $\alpha$ and $c_2$, so we consider solution with simple choice of $\alpha$, $m$ and $c_2$.

\subsubsection{Solution for $\alpha=4$, $m=-\frac{1}{2}$, but $c_2 \neq 0$:}
The equation \eqref{g2.18} with a simple choice of $\alpha=4$, $m=-\frac{1}{2}$ and $c_2 \neq 0$  reduces to
\be\label{g2.100a}
 {\Big(\frac{da}{d\hat{\tau}}\Big)}^2= 1-h_0^2 a^2(\hat{\tau})-\frac{H_0^2}{a^2(\hat{\tau})},
\ee
where $\hat{\tau}=\sqrt{c_2}\tau$, $h_0^2= -\frac{2u_0K^2}{c_2}$ and $H_0^2=- \frac{2v_0K^2}{c_2}$. Similar equation is also obtained in a recent work \cite{R2:bm} on $R^2$ gravity. The term  $\frac{H_0^2}{a^2(\hat{\tau})}$ in \eqref{g2.100a} is dominating in the early era of evolution, while $h_0^2a^2(\hat{\tau})$ has a large contribution in the later epoch. So the different terms in \eqref{g2.100a}  lead to distinct cosmic evolution and the evolution also depend on the signature of these terms. Accordingly  we consider the following piece-wise solutions which are relevant to the different domains of $a(\hat{\tau})$ according to greater contribution of individual terms in the right side of \eqref{g2.100a}.
Further from the field equations \eqref{g2.2} and \eqref{g2.3} with \eqref{g2.100a} NEC gives
\be\label{g2.100a1}
\rho_{_{eff}}+p_{_{eff}} =  \frac{2c_2}{K^2a^4}\Big(2H_0^2 -a^2  \Big),
\ee
so the NEC will be satisfied near the throat as long as $a< \sqrt{2}H_0$ and it is violated for $a> \sqrt{2}H_0$. Further the potential energy $V(\phi)$ as a function of $a(\tau)$ using \eqref{g2.100a}and \eqref{g2.10} is
\begin{equation}\label{g2.100a2}
V(a)= \frac{c_2}{2K^2}\Big(h_0^2-\frac{H_0^2}{a^4}\Big).
\end{equation}
In the late era at $ a^4> \frac{H_0^2}{h_0^2} $ the potential energy is positive, and asymptotically it leads to a constant $ \frac{c_2 h_0^2}{2K^2}$.
\subsubsection*{3.2.2A: Evolution in the domain when $ h_0^2 a^2(\hat{\tau})<<\frac{H_0^2}{a^2(\hat{\tau})}$ in the early universe:}
In the very early era near to the classical singularity contribution of $h_0^2 a^2(\hat{\tau})$ is very small compared to $\frac{H_0^2}{a^2(\hat{\tau})}$; so in the early era we can neglect the contribution of $h_0^2 a^2(\hat{\tau})$. Thus  \eqref{g2.100a} in the early era gives
\begin{equation}\label{g2.100b}
 a^2(\hat{\tau})=H_0^2 + \hat{\tau}^2
\end{equation}
apart from an integration constant. The equation \eqref{g2.100b} represents a wormhole solution and the radius at the throat is $H_0$. The scale factor is symmetric and increases with $ \hat{\tau} $. So, from \eqref{g2.100a1} the NEC is not violated as long as $\tau< H_0$.
 The dilaton field from \eqref{g2.9} and \eqref{g2.100b} gives
\begin{equation}\label{g2.100b1}
K~\phi=\pm \tan^{-1}(\frac{\hat{\tau}}{H_0})
\end{equation}
apart from a constant, which shows that $\phi=0$ at $\hat{\tau}=0$ and $\phi$ is real.
 Further at sufficiently large scale factor the contribution of $h_0^2 a^2(\hat{\tau})$ in \eqref{g2.100a} also influences evolution, so we consider contribution of all terms in \eqref{g2.100a} in next subsection.

\subsubsection*{3.2.2B: Evolution in the domain when $ h_0^2 a^2(\hat{\tau})$ is not negligible with $\frac{H_0^2}{a^2(\hat{\tau})}$:}
In this era the solution of \eqref{g2.100a} is
\begin{equation}\label{g2.100f}
 a^2(\hat{\tau})=\frac{1}{2h_0^2}\Big[1+(1-4h_0^2H_0^2)^{\frac{1}{2}}\cos(2h_0\hat{\tau})\Big],
\end{equation}
where $1> 4h_0^2H_0^2 $. The scale factor $a(\hat{\tau})$ lies in the range $ a_{-}^2 < a^2(\hat{\tau}) < a_{+}^2$
and $ a_{\pm}^2=\frac{1}{2h_0^2}\Big[1 \pm (1-4h_0^2H_0^2)^{\frac{1}{2}}\Big]$ satisfying the condition $h_0^2a^4(\hat{\tau}) +H_0^2 > a^2(\hat{\tau})$ in \eqref{g2.100a}. So, we have oscillation of the scale factor in Euclidean time. Now with analytic continuation with $\tau=it$, \eqref{g2.100f} gives\\
$ a^2(\hat{t})=\frac{1}{2h_0^2}\Big[1+(1-4h_0^2H_0^2)^{\frac{1}{2}}\cosh(2h_0\hat{t}~)\Big]$, which is a wormhole configuration with $t$ and asymptotically evolves exponentially.
\subsubsection*{3.2.2C: Evolution in the domain when $ h_0^2 a^2(\hat{\tau})>>\frac{H_0^2}{a^2(\hat{\tau})}$ at late era:}
In the later era $\frac{H_0^2}{a^2(\hat{\tau})} $ is negligible with the term $h_0^2a^2(\hat{\tau})$, so assuming  $\frac{H_0^2}{a^2(\hat{\tau})}\approx 0 $ in \eqref{g2.100a} the scale factor evolves as
\begin{equation}\label{g2.100g}
a^2(\hat{\tau})=\frac{1}{h_0^2}\sin^2(h_0 \hat{\tau})
\end{equation}
in the domain  $a(\hat{\tau})h_0< 1$, while in the domain $a(\hat{\tau})h_0> 1$
it gives
\begin{equation}\label{g2.100h}
 a(\hat{\tau})=\frac{1}{h_0}\cos(h_0 \hat{\tau}), ~~ \hbox{so with } \tau=i t, ~~a(\hat{t} )= \frac{1}{h_0}\cosh(h_0 \hat{t} ),
\end{equation}
apart from an integration constant. Further the potential energy from \eqref{g2.100a2} in this asymptotic domain is $ V= \frac{c_0h_0^2}{2K^2}$. This gives an expanding universe with cosmic time $t$ (where $\hat{t}= \sqrt{c_2}t$) and asymptotically at $h_0\hat{t}>>1$, the expansion is exponential.  Now the dilaton field from \eqref{g2.100h} and  \eqref{g2.9} gives
\begin{equation}\label{g2.100g1}
i K~\phi=\pm h_0 \hat{\tau}
\end{equation}
apart from an integration constant. So the scalar field $\phi$ is imaginary with respect to $\tau$. Now one can interpret the cosmic scenario of \eqref{g2.100h} as the solutions  of an oscillating universe  in the Euclidean time before crossing the deSitter radius $\frac{1}{h_0}$ and eventually the universe expands exponentially with proper time $t$ after crossing the deSitter radius. The constraint on the parameters in \eqref{g2.100f} gives $\frac{1}{h_0}> 2H_0$, which shows that the deSitter radius is always greater than the  radius of the wormhole at the throat in \eqref{g2.100b}.
\par
We have piece-wise solutions of \eqref{g2.100a} as \eqref{g2.100b}, \eqref{g2.100f}, \eqref{g2.100g} and \eqref{g2.100h} according to dominating contribution of different terms in the right side of \eqref{g2.100a}. In a nutshell we can interpret and unify these piece-wise solutions representing cosmic scenario of an universe beginning from the solution \eqref{g2.100b} in the very early era and subsequent evolution passes through era described by \eqref{g2.100f}. In solution \eqref{g2.100f} the universe oscillates with $\hat{\tau}$ after passing an expansion through  \eqref{g2.100b} in $\hat{\tau}$.  Finally the universe emerges to an expanding era with cosmic time $\hat{t}$ given by \eqref{g2.100h} after crossing the radius $\frac{1}{h_0}$ of an oscillating universe. Asymptotically we can achieve exponential expansion. Analytical solution of the field equations in general is quite non-trivial, so we consider numerical solution for $\kappa=1$, $\gamma=-1$ with a few standard potentials.
\section{Numerical solution of the Euclidean field equations with $V(\phi)$ and dilaton coupling $\Lambda(\phi)$:}
We present numerical solution of the field equations to study wormhole configuration assuming $\Lambda(\phi)$ and $V(\phi)$. We consider dilaton coupling $\Lambda(\phi)=\Lambda_0 e^{-\nu \phi}$ \cite{1:nep} in  the numerical solution, where $\Lambda_0$ and $\nu$ are constants.  The numerical solution using all the equations (2)-(4) gives us wormhole solutions for exponential and inverse power law potentials. Numerical solutions are obtained using initial conditions on $\Big(a(\tau), a'(\tau), \phi(\tau), \phi'(\tau)\Big)$. They yield two distinct categories of wormhole  depending on the potential. In one category we have usual model of wormhole of single lower bound, while in other models we have wormhole with multiple minima and maxima around a global minimum. The potential decays asymptotically with decaying amplitude in all cases.

\subsection{Solution with exponential potential $V_0 e^{-\mu\phi}$ and inverse power law potential $ V_0\phi^{-\alpha}$ for $\alpha= 2, 1, \frac{1}{2}$, etc. :}
The numerical solution using all the Euclidean field equations are presented graphically as  $a(\tau)$ versus $\tau$ for different potentials. The evolution of the scale factor and corresponding potential are shown respectively  in fig.1a and fig.1b for $V_0 e^{-\mu \phi}$ potential. Further the scale factors and the corresponding potentials are given respectively in fig.2a and fig.2b for inverse potential $V_0 \phi^{-\alpha}$ \cite{ratra:inverse} for $\alpha=2, 1, \frac{3}{2}$, etc.   The fig.2a is a superposition of plot of $a(\tau)$ with $\tau$ for $V_0\phi^{-2}$, $V_0\phi^{-1}$ and $V_0\phi^{-\frac{3}{2}}$ potentials considering three initial conditions for each of them.
The scale factor $a(\tau)$ is non-vanishing with a lower bound in each solution. In fig.1a, $a(\tau)$ increases with $\vert\tau\vert$ without any oscillations and $a(\tau)$ shows a single minimum for exponential potential, while in fig.2a there is an oscillation of $a(\tau)$ with $\tau$. In fig.2a,  $a(\tau)$ shows an overall ascending nature with increasing $\vert\tau\vert$ accompanying with multiple non-vanishing minima and finite maxima around a global non-vanishing minimum. The fluctuations of $a(\tau)$ with $\tau$ in fig.2a continues even for large $\tau$.  The plot of $a(\tau)$ versus $\tau$ is almost symmetric about the global minimum (or single minimum ) in each solution.
The scale factor $a(\tau)$ is finite with $\tau$ in all cases. So the solution of $a(\tau)$ versus $\tau$ represents wormhole configuration.
\par
The variation of potential $V(\phi)$ with $\tau$ are shown in fig.1b and fig.2b respectively for $V_0 e^{-\mu \phi}$ and $V_0 \phi^{-\alpha}$  potentials.  In all cases, the potential $V(\phi)$ is maximum near the throat of the wormhole and decays sharply without any oscillations from its maximum on both sides about this peak and asymptotically decays  significantly to very small, but finite value even at large $\tau$. The interpretation of above solutions are considered later on after presenting some other solutions with inverse power law potentials.
\par
Above characteristics of $a(\tau)$ with $\tau$ in fig.2a are almost identical with other class of inverse power law potentials with $\alpha= 2.6, \frac{5}{2•}, \frac{7}{3•}, \frac{5•}{3•},$ etc. The variation of $V(\phi)$ with $\tau $ are also identical for other power law potentials. Numerical solutions (see Appendix-I) of them are obtained using same initial condition for each $\alpha$, which are given in fig.7a and fig.7b respectively. An interesting feature is that expansion rate of $a(\tau)$ is greater for larger $\alpha$.

\begin{figure}
\centering
\begin{subfigure}{.45\textwidth}
\includegraphics[width=\linewidth]{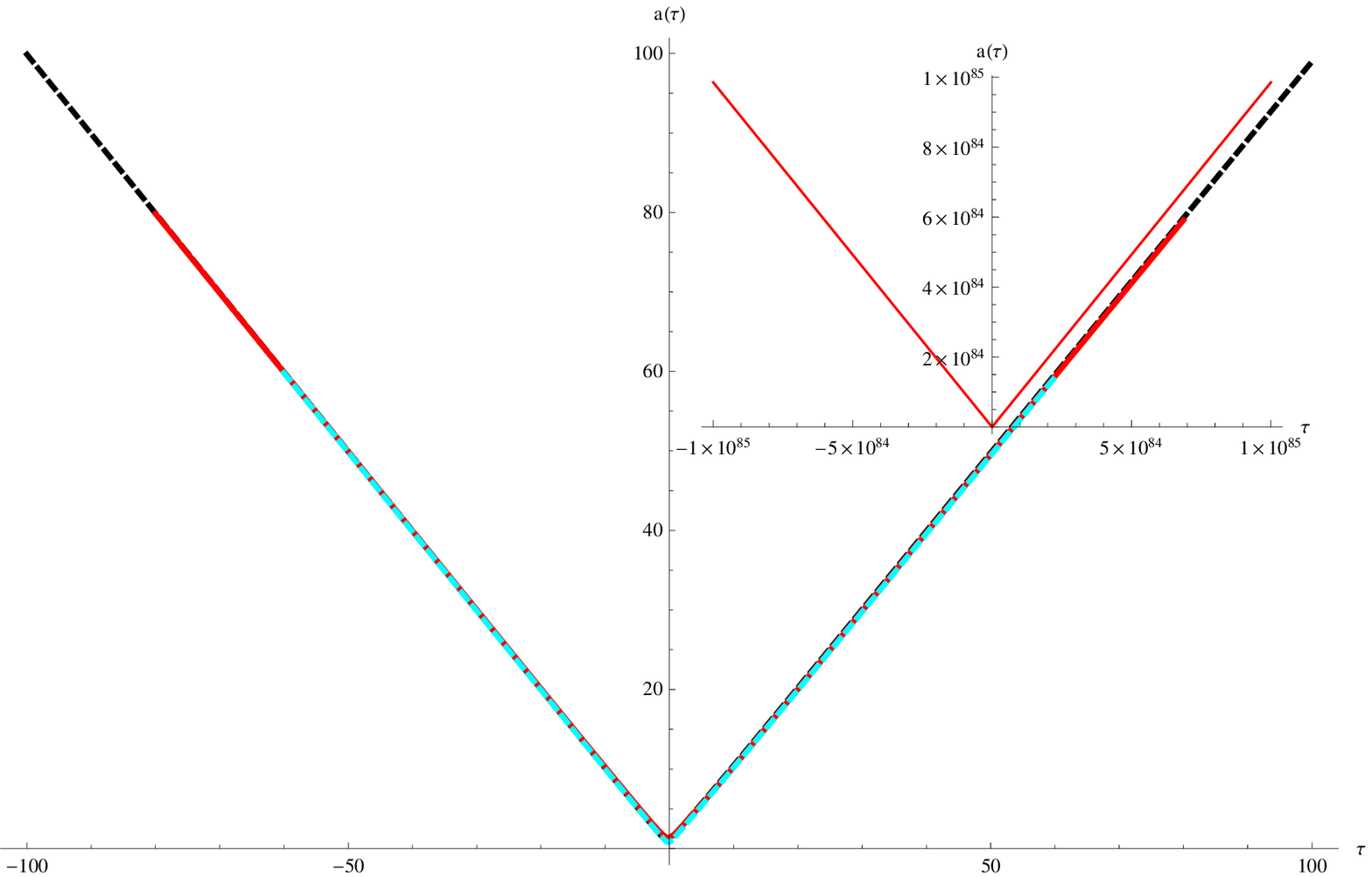}
\caption{fig.1a: $a(\tau)$ versus $\tau$}
\label{fig:sub1}
\end{subfigure}
\begin{subfigure}{.45\textwidth}
\includegraphics[width=\linewidth]{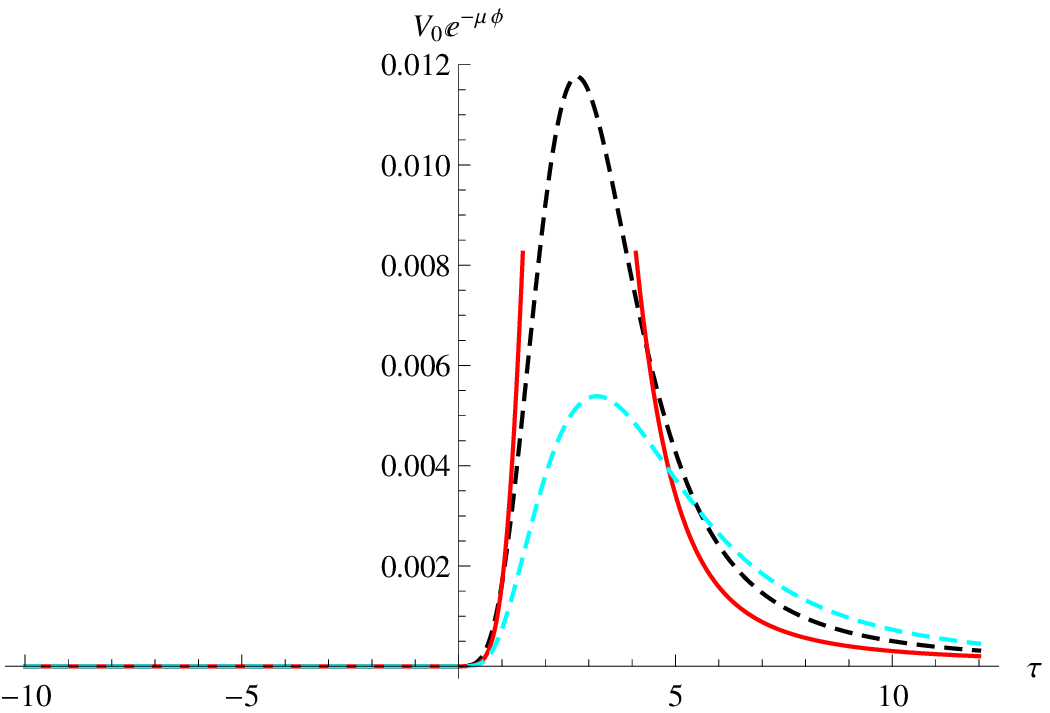}
\caption{fig.1b: $V(\phi)$ versus $\tau$}
\label{fig:sub2}
\end{subfigure}
\caption{The evolution of $a(\tau)$ and $V(\phi)$ with $\tau$ are shown respectively in fig.1a and fig.1b for $e^{-\mu\phi}$ potential with $\gamma=-1$, $V_0=1$, $K=1$, $\kappa=1$, $\nu=1$, $\mu=8$ and $\Lambda_0=1$. The initial conditions on $\Big(a(\tau),~a'(\tau),~\phi(\tau),~\phi'(\tau)\Big)$ for the solution along the black dashed,  red and cyan dashed curves are respectively $(1.9,
0.9, 0.8,-0.4)$, $(2, 0.8, 0.8, -.55)$ and $ (1.6, 0.9, 0.9, -0.4) $ at $\tau=1$.}
\label{figure 2}
\end{figure}

\begin{figure}
\centering
\begin{subfigure}{.45\textwidth}
  \includegraphics[width=\linewidth]{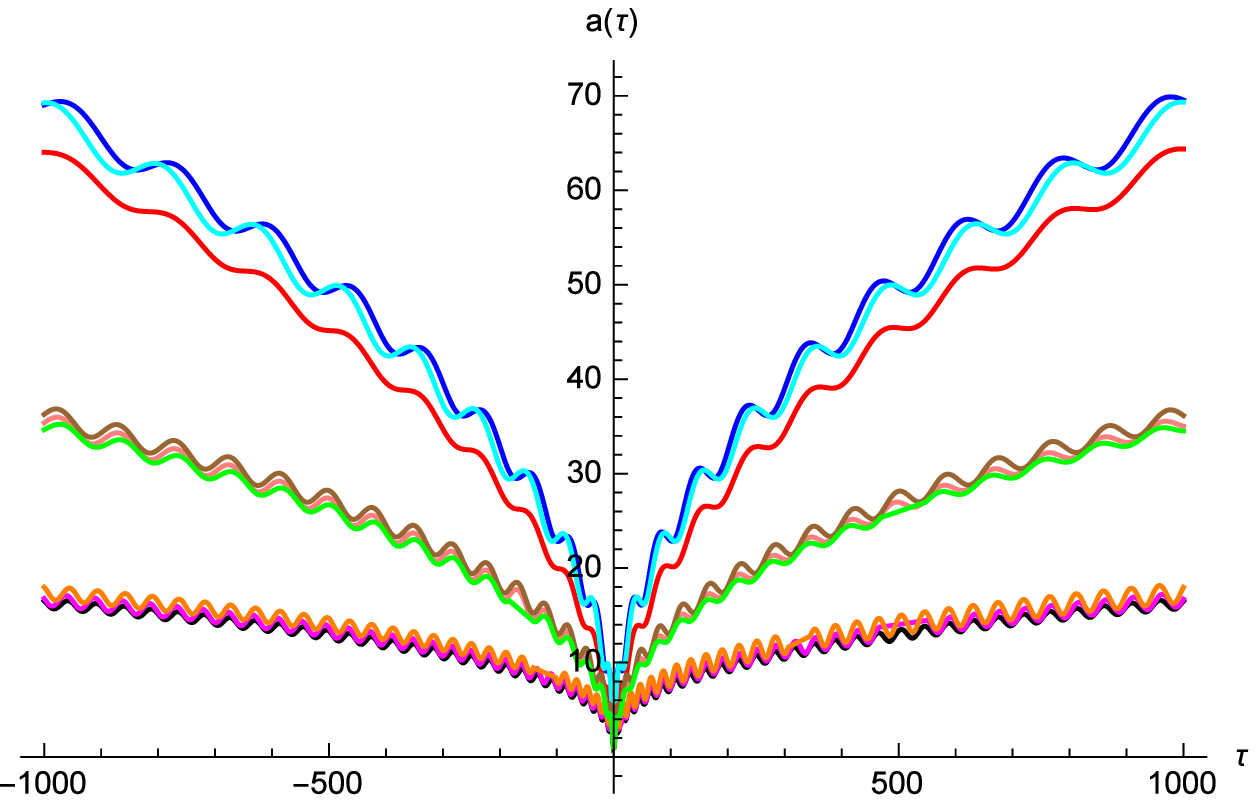}
  \caption{fig.2a: $a(\tau)$ versus $\tau$}
  \label{fig:sub1}
\end{subfigure}
\begin{subfigure}{.45\textwidth}
  \includegraphics[width=\linewidth]{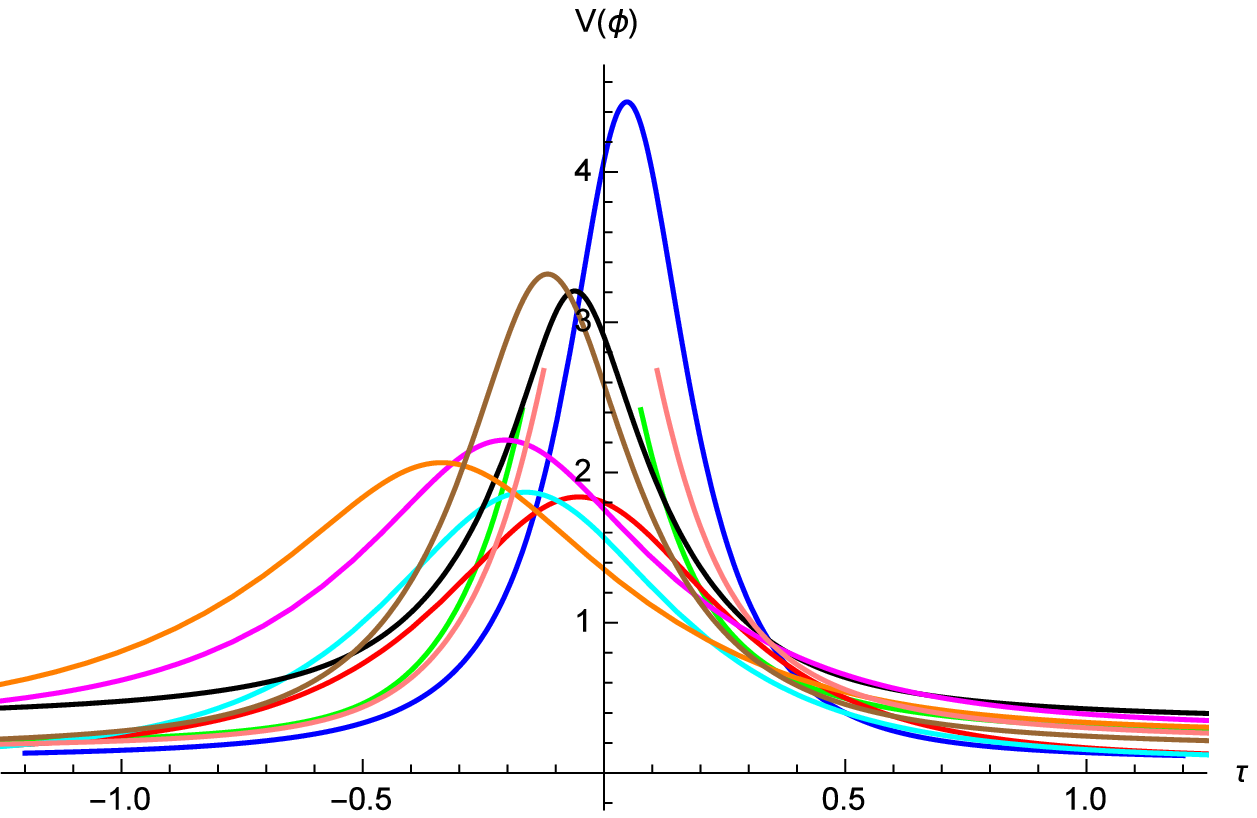}
 \caption{fig.2b: $V(\phi)$ versus $\tau$}
 \label{fig:sub2}
\end{subfigure}
\caption{The evolution of $a(\tau)$ and $V(\phi)$ with $\tau$ are shown respectively in fig.2a and fig.2b for inverse potential with $\gamma=-1$, $V_0=1$, $K=1$, $\kappa=1$, $\nu=0.7$ and $\Lambda_0=1$ . Three sets of initial conditions are chosen on $\Big(a(\tau),~a'(\tau),~\phi(\tau),~\phi'(\tau)\Big)$ at $\tau=0.1$ for each potential. The initial conditions on blue, red and cyan curves are respectively $(2, -0.512559, .5, 1)$, $(1.6, -0.4, 0.8, .8)$ and $( 2.5, -0.5, 0.9, 1.2)$ for $\phi^{-2}$ potential. The initial conditions on brown, green and pink curves are respectively $(1.5, -0.4, 0.7, 2)$, $( 1, -0.3, 0.61, 2.4)$ and $(1.2, -0.2, 0.5, 1.8)$  for $\phi^{-\frac{3}{2•}}$ potential, while black, magenta and orange curves are drawn with initial conditions $(1, -0.5, 0.5, 2)$, $(1.4, -0.5, 0.7, 1.5)$ and $(1.6, -0.7, 0.9, 1.8)$ for $\phi^{-1}$ potential. }
\label{figure 4}
\end{figure}

\begin{figure}
\centering
\begin{subfigure}{.4\textwidth}
  \includegraphics[width=\linewidth]{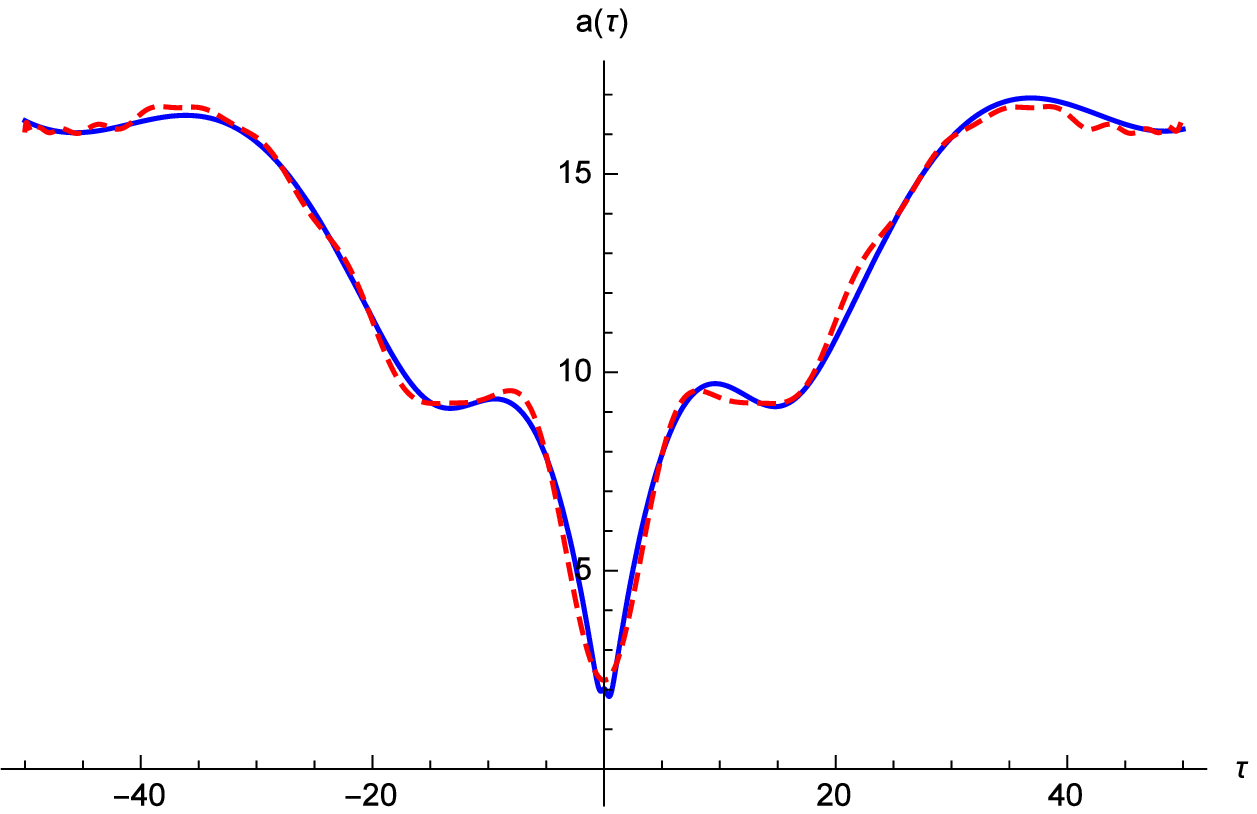}
  \caption{fig.3a: Plot of $a(\tau)$ (blue curve)\\ and corresponding fitted dashed magenta\\ curve using fit with polynomial of ``odd\\ and even" power of $\tau$ for $\phi^{-2}$ potential.}
  \label{fig:sub1}
\end{subfigure}
\begin{subfigure}{.4\textwidth}
\includegraphics[width=\linewidth]{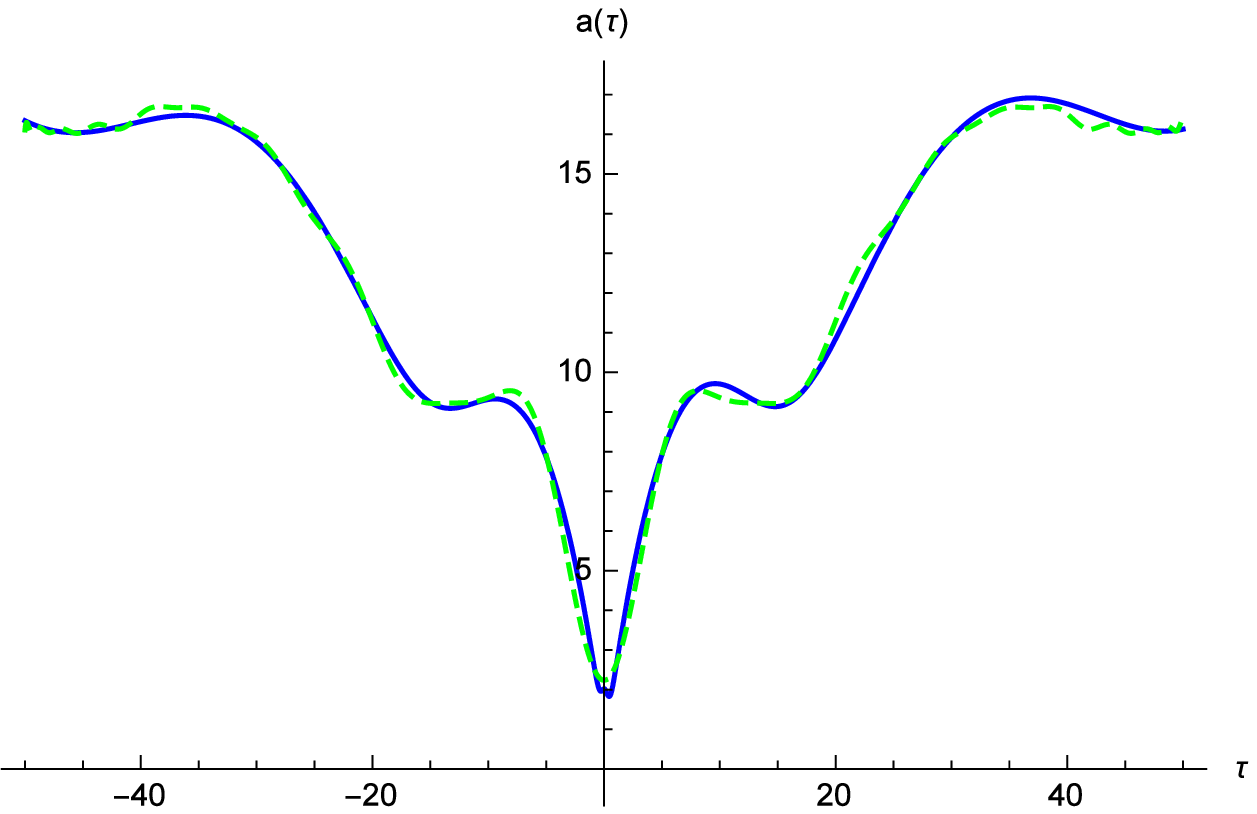}
 \caption{fig.3b: Plot of $a(\tau)$(blue curve) and corresponding fitted dashed green curve using fit with polynomial of `` even" power of $\tau$ for $\phi^{-2}$ potential.}
 \label{fig:sub2}
\end{subfigure}
\caption{The initial conditions in numerical solution of $a(\tau)$ shown in blue curve of fig.3a and fig.3b  are  $\Big(a(0.1)=2,~a'(0.1)=-0.5125559,~\phi(0.1)=0.5,~\phi'(0.1)=1\Big)$ with $\gamma=-1$, $V_0=1$, $K=1$, $\kappa=1$, $\nu=0.7$ and $\Lambda_0=1$.}
\label{figure 7}
\end{figure}

\begin{figure}
\centering
\begin{subfigure}{.45\textwidth}
  \includegraphics[width=\linewidth]{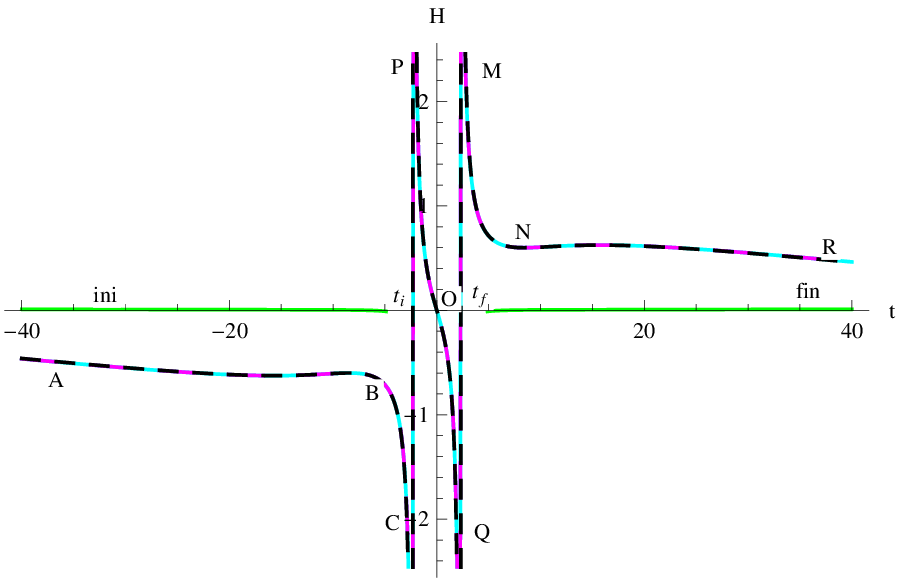}
  \caption{fig.4a: $H(t)$ versus $t$ for $\phi^{-2}$ potential}
  \label{fig:sub1}
\end{subfigure}
\begin{subfigure}{.45\textwidth}
  \includegraphics[width=\linewidth]{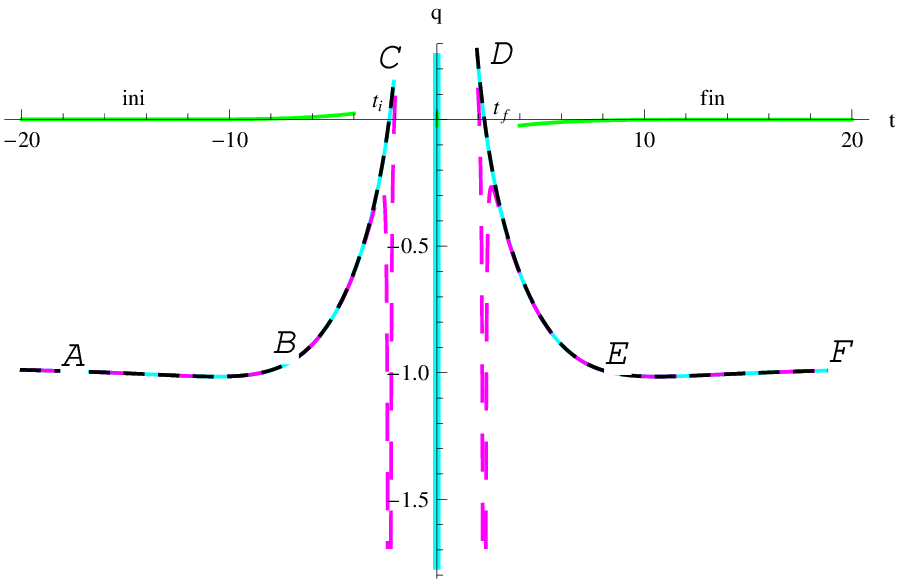}
 \caption{fig.4b: $q(t)$ versus $t$ for $\phi^{-2}$ potential}
 \label{fig:sub2}
\end{subfigure}
\caption{The evolution of Hubble parameter $H(t)$ and deceleration parameter $q(t)$ with $t$ are shown respectively in fig.4a and fig.4b. Real and imaginary parts of $H(t)$ and $q(t)$ obtained from the fit of $a(\tau)$ with polynomial of ``odd and even" power of $\tau$ are shown respectively in cyan and green curves. Again fit of $a(\tau)$ with polynomial of ``even" power of $\tau$ leads to real $H(t)$ and $q(t)$, which are shown by black dashed curves. The magenta dashed curves for $H(t)$ and $q(t)$ are plotted from the absolute value of a(t) using (36).}
\label{figure 8}
\end{figure}

\begin{figure}
\centering
\begin{subfigure}{.49\textwidth}
  \includegraphics[width=\linewidth]{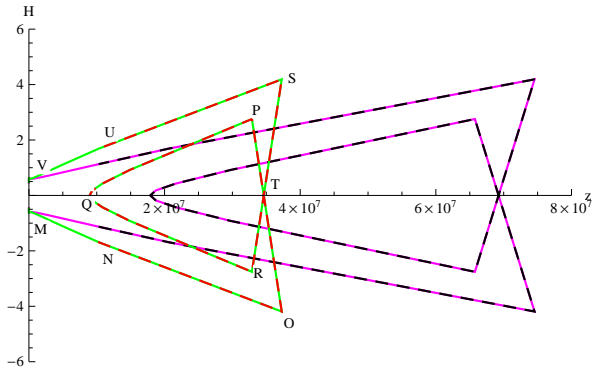}
  \caption{fig.5a: $H(z)$ versus $z$ for $\phi^{-2}$ potential}
  \label{fig:sub1}
\end{subfigure}
\begin{subfigure}{.49\textwidth}
  \includegraphics[width=\linewidth]{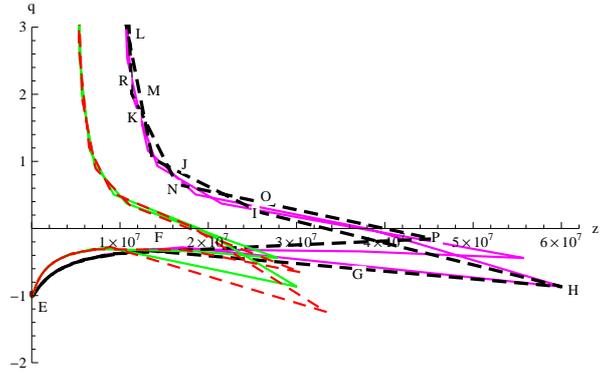}
 \caption{fig.5b: $q(z)$ versus $z$ for $\phi^{-2}$ potential}
 \label{fig:sub2}
\end{subfigure}
\caption{The plot of Hubble parameter $H(z)$ and deceleration parameter $q(z)$ with $z$ are shown respectively in fig.5a and fig.5b from the absolute value of $a(t)$ from $(36)$. The green curve and red dashed curve for $H(z)$ in fig.5a are plotted respectively within range $-38<t<38$ and $-3<t<3$ assuming $a_{0}=a_{abs}(t=27)=2\times10^{7}$; while the magenta curve and black dashed curve for $H(z)$ are obtained within range $-40<t<40$ and $-3.5<t<3.5$ respectively assuming $a_{0}=a_{abs}(t=29)=4\times10^{7}$. Further, the green curve and red dashed curve for $q(z)$ in fig.5b are shown respectively within range $-38.1<t<38.1$ and $-38.22<t<38.22$ assuming $a_{0}=a_{abs}(t=26)=1\times10^{7}$; while the magenta curve and black dashed curve for $q(z)$ are obtained within range $-38.2<t<38.2$ and $-38.4<t<38.4$ respectively assuming $a_{0}=a_{abs}(t=27)=2\times10^{7}$.}
\label{figure 8}
\end{figure}

\begin{figure}
\centering
\includegraphics[width=10cm,height=6cm]{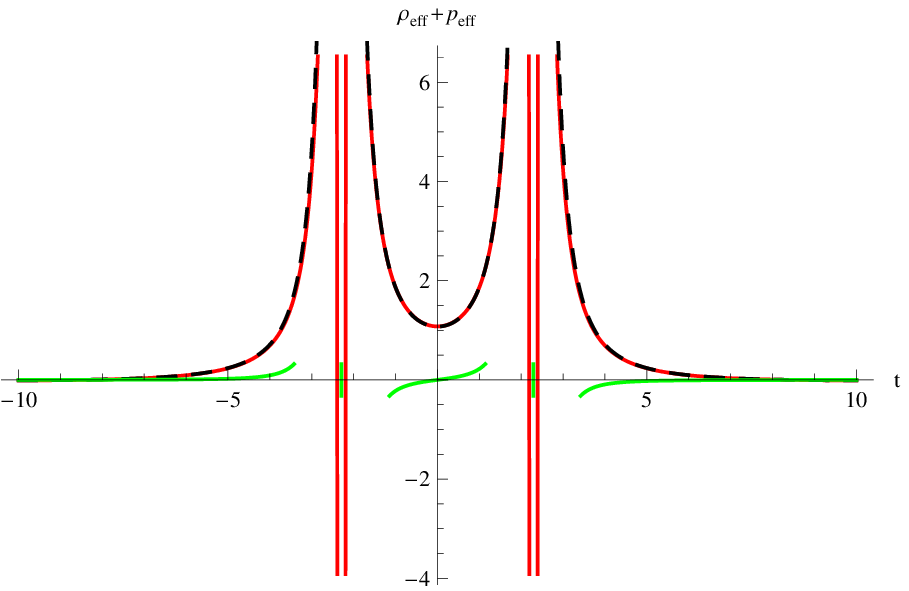}
  \caption{The evolution of $\rho_{eff}+p_{eff}$ with $t$ for $\phi^{-2}$ potential. The red and green curves are shown respectively for real and imaginary parts of $\rho_{eff}+p_{eff}$ using fit of $a(\tau)$ by polynomial of odd and even power of $\tau$ given in \eqref{g2.42}, while black curve is obtained by fit of $a(\tau)$ using polynomial of even power of $\tau$ given in \eqref{g2.43}.}
 \label{figure 5}
\end{figure}

\section{Cosmic scenario of the early universe from curve fit of the numerical solution:}
Now, we explain dynamical consequence of above solutions and hence functional form of the scale factor is necessary to obtain cosmic evolution of the early universe. So, we consider curve fitting of the numerical solution of $a(\tau)$ close to the throat of wormhole for $V_0\phi^{-2}$ potential. We also consider curve fitting for exponential potential (given in Appendix III). The Hubble expansion $H(t)$ and deceleration parameter $q(t)$ \big(as well as $H(z)$ and $q(z)$\big) are the relevant parameters at first-hand to determine cosmic evolution and we can evaluate them by using expression of $a(t)$. The scale factor $a(t)$ is evaluated from $a(\tau)$, which is obtained by fit of $a(\tau)$ from the numerical solution and analytic continuation $\tau=it$. The absolute values of $a(t)$ also yield the parameter $H(z)$ and $q(z)$. The evolution of $H(t)$ and $q(t)$ are almost independent of potentials. The plot of $H$ and $q$ as a function of $t$ yield cosmic evolution from above fitting. Two distinct expressions of $a(\tau)$ may appear depending on the curve fitting in each wormhole which are given in the following section. The explicit expressions of $a(\tau)$ are presented in Appendix 7.2 in equation (36) and (37).
\subsection{Cosmic evolution near the throat of wormhole by curve fit of $a(\tau)$ with a polynomial of $\tau$ for $V_0\phi^{-2}$ potentials:}
To interpret the numerical solution we consider curve fitting of $a(\tau)$ within a small domain about the throat of wormhole. Two distinct expressions of $a(\tau)$ may be obtained depending on the fit either with a polynomial of ``odd and even" power of $\tau$, or with a polynomial of ``even" power of $\tau$ in each numerical solution. Those fitted curves are shown in fig.3a and fig.3b respectively for polynomial of ``odd and even" and ``even" power of $\tau$ for  $V_0\phi^{-2}$ potential. The form  of $a(\tau)$ considering  polynomial of ``odd and even" power of $\tau$ reads as
\begin{equation}\label{g2.38}
a(\tau)= \sum_{_{n=0}}^{N} b_n \tau^n,
\end{equation}
while for polynomial of ``even" power of $\tau$ is
\begin{equation}\label{g2.39}
a(\tau)= \sum_{_{n=0}}^{M} c_{2n} \tau^{2n},
\end{equation}
where $b_n$ and $c_{2n}$ are the coefficients in the polynomials, and they depend on the potential and initial conditions. The explicit form of \eqref{g2.38} and \eqref{g2.39} for the potential $V_0\phi^{-2}$ are given respectively in \eqref{g2.42} and \eqref{g2.43} in the Appendix-II for detailed information. We assume the functions $1, \tau, \tau^2, \tau^3, \tau^4, .....$ with $N=50$ in \eqref{g2.38} for $\phi^{-2}$ potential in the curve fit of numerical solution. Further we choose $M=25$ in the fit with even power of $\tau$ in equation \eqref{g2.39}.  The fitted curves are shown in fig.8a and fig.8b respectively for polynomial of ``odd and even" and ``even" power of $\tau$ for  $V_0 e^{-\mu\phi}$ potential. Again the expression of \eqref{g2.38} and \eqref{g2.39} for $V_0 e^{-\mu\phi}$ potential are given respectively in \eqref{g2.44} and \eqref{g2.45} in the Appendix-III.
\par
The fit with only even or odd powers of $\tau$ does not reveal good fit, however a fit with all polynomial of $\tau$ in \eqref{g2.38} yields a better one. Hence from  the fit the metric tensor in Euclidean space $g_{ik} \propto a^2(\tau) \delta_{ik}$ are determined as a function of $\tau$ and $g_{00}=1  $; where $i, k$ runs $1, 2, 3$. Analytic continuation in the Euclidean space by  $\tau=it$ gives the metric tensor in terms of time $t$. However, the new  spacetime appears  to be distinct  from a Lorentz spacetime due to presence of real and imaginary parts of $a(t)$. The odd power of $\tau$ in the fit gives rise imaginary parts of $a(t)$ and the parameters dependent on $a(t)$ also give rise  real and imaginary parts. This is not surprising, as quantum mechanical process is dominant near the throat. In fact a wormhole configuration in the  domain ($ t_i \leq t \leq t_f$) around the throat shows unusual evolution, which is analogous to the classical forbidden domain in the analytic
solution in \cite{R:Comm} ($t_i$ and $t_f$ are specified  in the next section). The coordinate ``$t$" in the classical
allowed domain is the usual time, while ``$t$" in the classical forbidden domain changes its usual notion of cosmic time, and the observables may have both real and imaginary parts in some domain. However, we can alleviate the imaginary parts from the observables using a fit of $a(\tau)$ with even power of $\tau$.
\subsection{Evolution of observable parameters $H(t)$ and $q(t)$ from the fit and interpretation:}
The parameters  $H(t)= \frac{\dot a}{a}$ and  $q(t)= -\frac{a \ddot{a}}{{\dot{a}}^2}$ are evaluated both from \eqref{g2.42} and \eqref{g2.43}, where $\dot{a}= \frac{da(t)}{dt}$. The plot of $H(t)$ and $q(t)$ are shown respectively in fig.4a and fig.4b for $\phi^{-2}$ potential, while for $e^{-\mu\phi}$ potential, $H(t)$ and $q(t)$ are shown respectively in fig.9a and fig.9b. In  above figures the initial collapsing era ends at $t=t_i$, while the final expansion initiated at $t=t_f$.The evolution of $H(t)$ and $q(t)$ in above solution are distinct from the usual cosmic evolution in cosmology. However in a recent work \cite{R:Comm}, we get analogues  evolution of $H(t)$ both from analytic and numerical solutions, but with distinct action.
\par
The parameters $H(t)$ and $q(t)$ have both real (shown by cyan curves) and imaginary (drawn by green curves) parts close to the throat, when $a(\tau)$  is fitted with polynomial of ``odd and even" power of $\tau$. Again, $H(t)$ and $q(t)$ obtained by using fit with polynomial of ``even" power of $\tau$ yields real value ( black dashed  curves in fig.4a and fig.4b), which are identical with the respective real parts of $H(t)$ and $q(t)$ obtained  with fit of ``odd and even" power of $\tau$. We further using the absolute value of $a(t)$ from (36) determine $H(t)$ and $q(t)$  which are presented by magenta curves in fig.4. $H(t)$ denoted by magenta curve is identical with above real part (cyan curve) of $H(t)$, whereas $q(t)$ denoted by magenta curve coincides with the real (cyan) values of $q(t)$ in the asymptotic region. Real parts of them dominates over the imaginary parts away from the throat.  Imaginary parts of them almost vanishes at large $t$, and asymptotically $H(t)$ approaches to a constant value and  parameter $q(t) \rightarrow -1$ simultaneously away from the throat. Thus an inflationary era can be realised  from the wormhole solution by analytic continuation $\tau=it$ at time $t$ away from the throat irrespective of the curve fit. The evolution of $H(t)$ and $q(t)$ for $\phi^{-2}$ potential are identical with those $H(t)$ and $q(t)$ for $e^{-\mu\phi}$ potential, except the value of $t_i$ and $t_f$.
\par
We further plot the Hubble parameter $H(z)$ and the deceleration parameter $q(z)$ as a function of red shift parameter $z$ in fig.5 from (36) using the absolute form of the scale factor $a_{abs}(t)$. The relation between the red shift parameter $z$ and the scale factor $a(t)$ is $1+z=\frac{a_{0}}{a_{e}}$, where $a_{e}$ is the scale factor at the time when the signal is emitted and $a_{0}$ is the scale factor at the time when the emitted signal is observed. Further, the value of $z$ is very large in the very early universe. Hence, the  parameter $z$ attains a large value in the radiation dominated era and $z$ is nearly equal to 3200 at matter-radiation equality domain. The Hubble parameter $H$ decreases with decrease of $z$ in the usual cosmic evolution; consequently $z$ is very small in the present late time universe. However, in the inflationary era or pre-inflationary era the value of $z$ is unknown. So, we have to set the value of $a_{0}$ in the expression of $1+z$ to evaluate $z$ in the very early universe. As we have no observational data, we use the plots of $H(t)$ versus $t$ in fig.4a to find $a_{0}$ knowing the absolute value of $a(t)$ from (36). It is observed from $H(t)$ versus $t$ in fig.4a that when $t>10$  \big(or $t<-10$ \big), $H(t)$ is almost constant in the asymptotic domain. So, we can logically choose the absolute value of $a(t)$ as $a_{0}$ in the domain $t>10$ \big( or $t<-10$ \big) using (36). In the plots of $H(z)$ versus $z$ \big(or $q(z)$ versus $z$\big) we assume the value of $a_{0}$ as $a_{0}=a_{abs}( t=26)=1\times10^{7}$; $a_{0}=a_{abs}(t=27)=2\times10^{7}$; $a_{0}=a_{abs}(t=29)=4\times10^{7}$. Evolution of $H(z)$ and $q(z)$ with $z$ are obtained using list plot. The plots of $H(t)$ and $q(t)$ with $t$ in fig.4 describe the evolution in both the domains which are symmetrically placed about t=0. Similarly, the plots of $H(z)$ and $q(z)$ with $z$ obtained from above consideration should also represent the evolution in both the domains. So, both the collapsing as well as expanding domains can be explained from the evolution of $H(z)$ \big(or $q(z)$\big) along with some unusual features about the classical singularity.
\par
It is difficult to interpret evolution of $H(z)$ \big(or $q(z)$\big) as $a_{abs}(t)$ is a polynomial of $t$ i.e. $z(t)$ is also a polynomial of $t$ and hence $H(z)$ and $q(z)$ both are multi-valued function of $z$. Different curves for $H(z)$ \big( or $q(z)$\big) appear due to different domains of $t$ used in the list plot of $H(z)$ and $q(z)$ as well as different choices of $a_{0}$ in the calculation of $z$. However, comparing $H(z)$ vs $z$ in fig.5a (or $q(z)$ vs $z$ in fig.5b) with $H(t)$ vs $t$ in fig.4a (or $q(t)$ vs $t$ in fig.4b), the chronological evolution of $H(z)$ with $z$  \big(or $q(z)$ with $z$ \big) in fig.5a and fig.5b can be obtained. It is observed from fig.4a that initially $H(t)$ is negative (i.e. $H(t)<0$ ) for $t<t_{i}$ and $H(t)$ vanishes at three points at $t=t_{i}$, $t=0$ and $t=t_{f}$ and finally $H(t)$ is positive ( i.e. $H(t)>0$ ) for $t>t_{f}$. The sequence of the points giving evolution of $H(t)$ with $t$ in fig.4a through points A, B, C, $t_{i}$, P, O, Q, $t_{f}$, M, N, R is equivalent to the evolution of $H(z)$ in fig.5a. The red dashed curve of $H(z)$ in fig.5a evolves through the points N, O, T, P, Q, R, S, U. The points T, Q and again T in $H(z)$ curve are equivalent to $t=t_{i}$, $t=0$ and $t=t_{f}$ respectively in $H(t)$ curve where $H(z)=0$. The sequence of points may also begin at M, pass through N, O, T, P, Q, R, S, U and finally terminate at V. The asymptotic values of $H(z)$ at the points M and V are identical with opposite sign for different values of $a_{0}$ and also for different domains of $t$. The constant asymptotic values of $H(z)$ at points M and V yield final inflationary era away from the throat similar to $H(t)$ versus $t$ curve in fig.4a. The $H(z)$ versus $z$ curves show identical nature with different values of $a_{0}$, but with increase of value of the choice $a_{0}$, the domain of $z$ in $H(z)$ curve increases. Further, when $H(z)$ is plotted in small domain of time interval ($-3<t<3$ in red dashed curve of fig.5a), then $H(z)$ assumes asymptotic value at points N, U at $z=1\times10^{7}$ near the throat of the wormhole; whereas for large domain of time interval ($-38<t<38$ in green curve of fig.5b), $H(z)$ assumes asymptotic value at points M,V at $z=0$ away from the throat of the wormhole.
\par
Futher, the nature of $q(z)$ curves for different $a_{0}$ are not identical; however evolution of $q(z)$ follows the same way of transition from negative to positive value and again positive to negative value \big(q(t) also shows identical evolution with $t$ in fig.4b \big). Further, $q(z)$ curve in black dashed curve in fig.5b beginning from the point E evolves either through the points F, G, H, I, J, K, M, or F, P, O, N, K, R and finally attains the same point L. The negative value of $q(z)$ at point E in fig.5b (which is equivalent to point A in $q(t)$ curve in fig.4b) after attaining the final point L returns back either along L, R, K, N, O, P, F or L, M, K, J, I, H, G, F and finally to E (which is equivalent to point F in $q(t)$ curve in fig.4b) which is asymptotic value of $q(z)$. All the curves of $q(z)$ begin initially from the same point E and asymptotically ends at the same point E irrespective of the values of $a_{0}$ and different domains of $t$ used in list plot. Thus, the asymptotic value of $q(z)$ at point E ($q(z)=-1$) in fig.5b  demonstrates the idea of final inflationary era similar to $q(t)$ in fig.4b far away from the throat of wormhole.
\par
Null energy condition is obtained from the field equations using solution \eqref{g2.42} and \eqref{g2.43} respectively for fit given in fig.3a and fig.3b. The plots of  $\rho_{_{eff}}+p_{_{eff}}$ versus $t$ are given in fig.6, which shows that the NEC is not violated in the neighbourhood of the throat. Further  $\rho_{_{eff}}+p_{_{eff}}$ vanishes far away from the throat. Imaginary part of  $\rho_{_{eff}}+p_{_{eff}}$ is very small with respect to its real part. The imaginary part appeared due to choice of fit of $a(\tau)$ with polynomial of both odd and even power of $\tau$. The imaginary part can be  removed using fit of $a(\tau)$ with polynomial of even power of $\tau$ and all the observables turned to be real.

\par
 It is  to note that $g_{ik}$ or $a^2(t)$ from \eqref{g2.38} or \eqref{g2.39} vanishes at $t= t_{i}$ and $t=t_{f}$, where $t_i=-2.7$ and $t_f= +2.7$ for $\phi^{-2}$ potential. Further evolution of $H(t)$ in the domain $t<t_i$ is a collapsing mode, while the domain $t>t_f$  is an expanding era in fig.4a. In between above collapsing and expanding modes, an unusual evolution is evident in the domain $t_i\leq t\leq t_f$ surrounding the throat with respect to $t$, wherein $a(\tau)$ is well behaved with $\tau$. In fact $t_i\leq t\leq t_f$ is the classical forbidden domain which is similar to the analytic result in \cite{R:Comm}; while the domains $t<t_i$ and $t>t_f$ are classical allowed with respect to time $t$.
\par
Imaginary parts of $H(t)$ and $q(t)$ obtained above are questionable from the observational ground, however, the imaginary part of the $H(t)$ may lead to a very small oscillation  of $a(t)$  near the throat.Again the potential is maximum near the throat and the fall of potential is evident from the plot of $V(\phi)$ with $\tau$.\\
The potential is  decaying, but it is vanishingly small even at large $\tau$. An estimate of decrease for $ V_0 e^{-\mu\phi}$ potential yields
$\frac{V(\tau=2.7)~~}{V(\tau=10^{30})}\approx 10^{63}$ and $\frac{V(\tau=2.7)~~}{V(\tau=10^{65})}\approx 10^{132}$ with initial condition given in black(dashed) curve of fig.1b. Thus the inflationary expansion is a consequence of initial wormhole configuration  in the dilaton Einstein Gauss-Bonnet theory.

\subsection{Possibility of multiple minima and maxima in a wormhole with inverse potential:}
The evolution of the scale factor $a(\tau)$ with $\tau$ for inverse potential shows multiple maxima and minima about the global minimum at the throat of wormhole.
Interestingly, multiple maxima and minima in above plot of $a(\tau)$ are not reflected in cosmic evolution of Hubble parameter $H(t)$, rather they represent an average nature \cite{inflation:H}. However, one may get hint  from expression of $a''(\tau)$, since $a''>0$ at the minima, while at the maxima $a''<0$.
Now from \eqref{g2.2} and \eqref{g2.3}, we have
\begin{equation}\label{g2.40}
a''(\tau)=-b(\tau) a'(\tau) -\Big(b'(\tau)+ \omega_0^2(\tau)\Big)a(\tau)
\end{equation}
where $b(\tau)= \frac{K^2}{2}\Lambda'(\frac{a'^2}{a^2•}-\frac{\kappa}{a^2})$,
 $\omega_0^2(\tau) = \frac{K^2}{3}\Big(\gamma \phi'^2 +V(\phi)\Big) $. The evolution of $a(\tau)$ from \eqref{g2.40} is determined by the functions of $b(\tau)$ and $\omega_0^2(\tau)$. The function of $b(\tau)$ determines the effect of damping in the evolution of $a(\tau)$, whereas the oscillation, if it exists at all, depends on both $b'(\tau)$ and $\omega_0^2(\tau)$. The damping is more effective for $b>0$, while for $b<0$ it may lead acceleration. Further $\omega_0^2$ may alter its sign  in a rapidly oscillating field $\phi$. Again at the extrema  of $a(\tau)$, the  equation \eqref{g2.40} gives
\begin{equation}\label{g2.41}
a''_0=\frac{K^2}{3}(-\gamma\phi'^2_0-V_0) a_0 +\frac{\kappa K^2}{2}\frac{\Lambda_0''}{a_0}= \frac{K^2}{3}(\phi'^2_0-V_0) a_0 +\frac{ K^2\Lambda_0}{2 a_0}(-\nu \phi''_0+ \nu^2\phi'^2_0)e^{-\nu\phi_0},
\end{equation}
for $\kappa=1$ and $\gamma=-1$.
The Gauss Bonnet term is assumed to be a small correction to the gravity, however the second term $\frac{\Lambda_0''}{a_0}$ in \eqref{g2.41} may be large in the early universe, while the first term in \eqref{g2.41} will be large at later era. The values of $a''(\tau)$ (i.e. either positive or negative) at the extrema are determined by the fluctuating field $\phi$, coupling function, and the potential at the extrema.
\section{Discussion:}
 Some analytic and numerical wormhole solutions are presented in the 4-dimensional Robertson Walker Euclidean background in the Einstein Gauss-Bonnet dilaton theory. Analytic solution in general is not trivial; so the solutions are obtained with simplified assumption for $\kappa=0$ with a restriction on the coupling function $\Lambda(\phi)$ as $\Lambda'\frac{a'}{a}=\frac{2m}{K^2}$. The problem of the cosmic singularity is avoided in the wormhole configuration with $\tau$.  Further, the wormhole in analytic solutions (subsection 3.2) transforms to an exponential expansion with $t$ using $\tau=it$ after crossing a phase of oscillating universe having a deSitter radius. In another analytic solution radiation (subsection 3.1.1) dominated era can be recovered  from a early wormhole configuration.

\par
We here present numerical solution of wormhole with exponential and a few inverse power law potentials with a plot of $a(\tau)$ versus $\tau$. Wormhole solutions with inverse power law potentials revealed multiple local maxima and minima about a global minimum at the throat unlike usual wormhole, while for exponential potential, we have usual wormhole of single minimum. The consequence of numerical solution of the wormholes are studied by curve fit of $a(\tau)$ for $\phi^{-2}$ potential. We evaluate $a(t)$ from fitted polynomial of $a(\tau)$ by  $\tau=it$. The expression of $a(t)$ is then used to evaluate the Hubble parameter $H(t)$ and the deceleration parameter $q(t)$ to study cosmic evolution. Consequently it leads to two distinct set of variables $\{ a(t), H(t),  q(t)\} $ depending on the fit of $a(\tau)$ either with polynomial of ``odd and even" or only ``even" power of $\tau$.
\par
The evolution of $H(t)$ (or real part of $H(t)$ or $H(t)$ obtained from the absolute form of $a(t)$ ) shows initial collapsing phase till $t<t_i$,  while $t> t_f$ is the final expanding  phase after evolving through the throat of wormhole and the parameter $H(t)$ (or real part of $H(t)$) shows unusual evolution within  $t_i \leq t \leq t_f$ around the throat. This domain  may be considered as classical forbidden regime. Above parameters  have both real and imaginary parts for fit of  $a(\tau)$ with polynomial of ``odd and even" power of $\tau$. Again the real parts of them dominate over the imaginary parts outside  domain $t_i \leq t \leq t_f$ and the imaginary parts of them vanish far away from the  domain. Imaginary parts of the parameters $\{ H(t), q(t)\}$ give rise oscillating $a(t)$, and also questionable from the observational point of view. However, we can alleviate the  imaginary parts of them using fit of $a(\tau)$ with polynomial of only ``even" power of $\tau$, though the fit of $a(\tau)$ with ``odd and even" power of $\tau$ is better one. The plot of $H(t)$ and $q(t)$  evaluated from the fit with polynomial of ``even" power of $\tau$ are identical with the real parts of $H(t)$ and $q(t)$ obtained from ``odd and even" power of $\tau$.
\par
In the asymptotic domain (i.e. $t>>t_f$) the Hubble parameter $H(t)$ \big(or $H(z)$\big) approaches to a constant value  and the deceleration parameter $q(t)$ \big(or $q(z)$\big) simultaneously approaches to $ -1$. Further, the potential $V(\phi)$ approaches to very small value at $t>>t_f$. An estimate of decrease of potential yields $\frac{V(\tau=2.7)~~}{V(\tau=10^{30})}\approx 10^{63}$ and $\frac{V(\tau=2.7)~~}{V(\tau=10^{65})}\approx 10^{132}$ for $ V_0 e^{-\mu\phi}$ potential with initial condition given in black curve of fig.1b. So the cosmic scenario of a Euclidean wormhole leads to an exponential expanding era under analytic continuation by $\tau=it$.  We have seen that these results are true for other standard potentials. Thus it appears that inflationary expansion is a consequence of the wormhole solution in the Euclidean space in the Einstein Gauss-Bonnet dilaton theory. This is the new feature in the literature. We further present wormhole solutions (Appendix-I) with some values of $\alpha$ in the inverse potential $\phi^{-\alpha}$.

\section{Appendix}
\subsection{ Appendix-I: Wormhole solution with different potentials but with  same initial condition:}
It is observed that the numerical solutions allow wormhole configuration  with inverse power law potential $\phi^{-\alpha}$ for $\alpha = \frac{5}{2•},~ \frac{7}{3•},~ 2,~ \frac{5•}{3•},$ etc under same initial condition for all. These solutions are presented in the fig.6.  It is observed from the numerical solutions that approximate highest value of index $\alpha$ is $\alpha=2.6$  in $\phi^{-\alpha}$ potential to get wormhole. The scale factor increases to a faster rate for higher values of $\alpha$.

\begin{figure}
\centering
\begin{subfigure}{.45\textwidth}
\includegraphics[width=\linewidth]{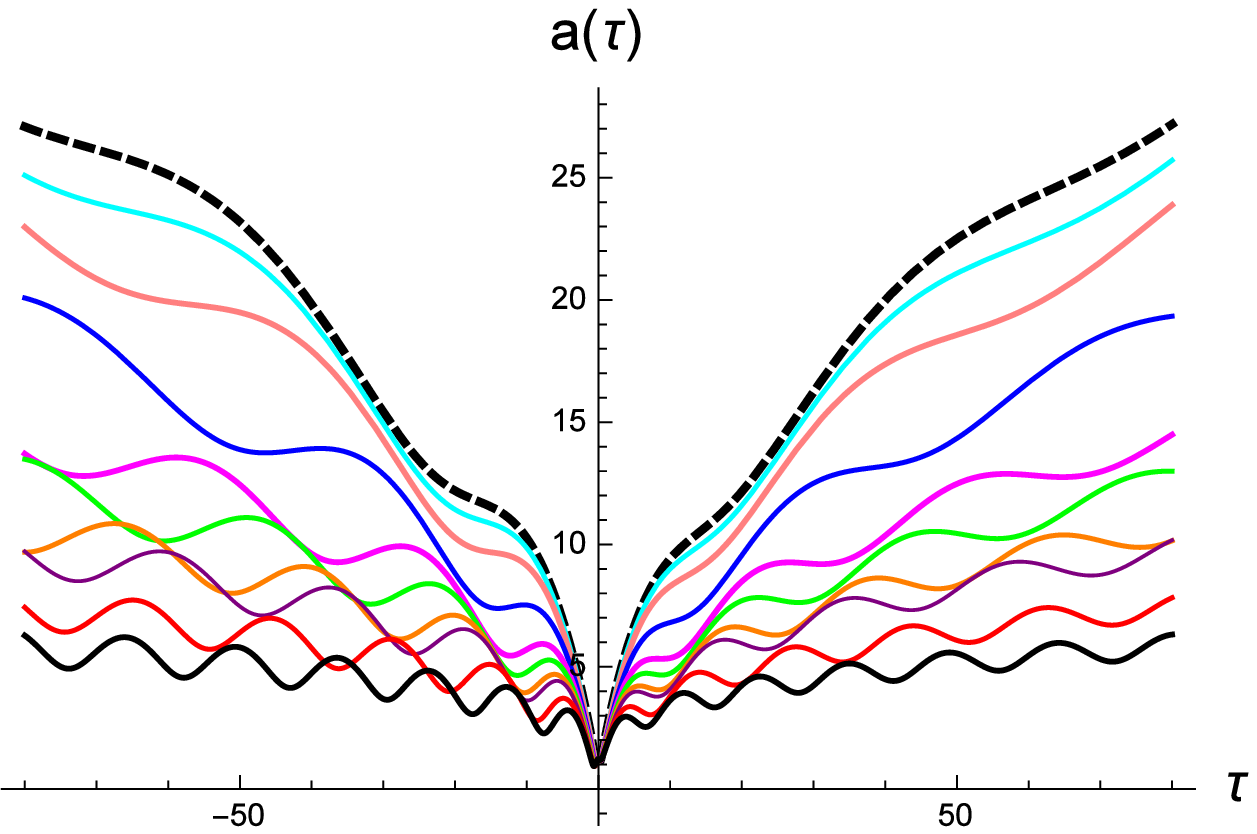}
\caption{fig.7a: $a(\tau)$ versus $\tau$}
\label{fig:sub1}
\end{subfigure}
\begin{subfigure}{.45\textwidth}
\includegraphics[width=\linewidth]{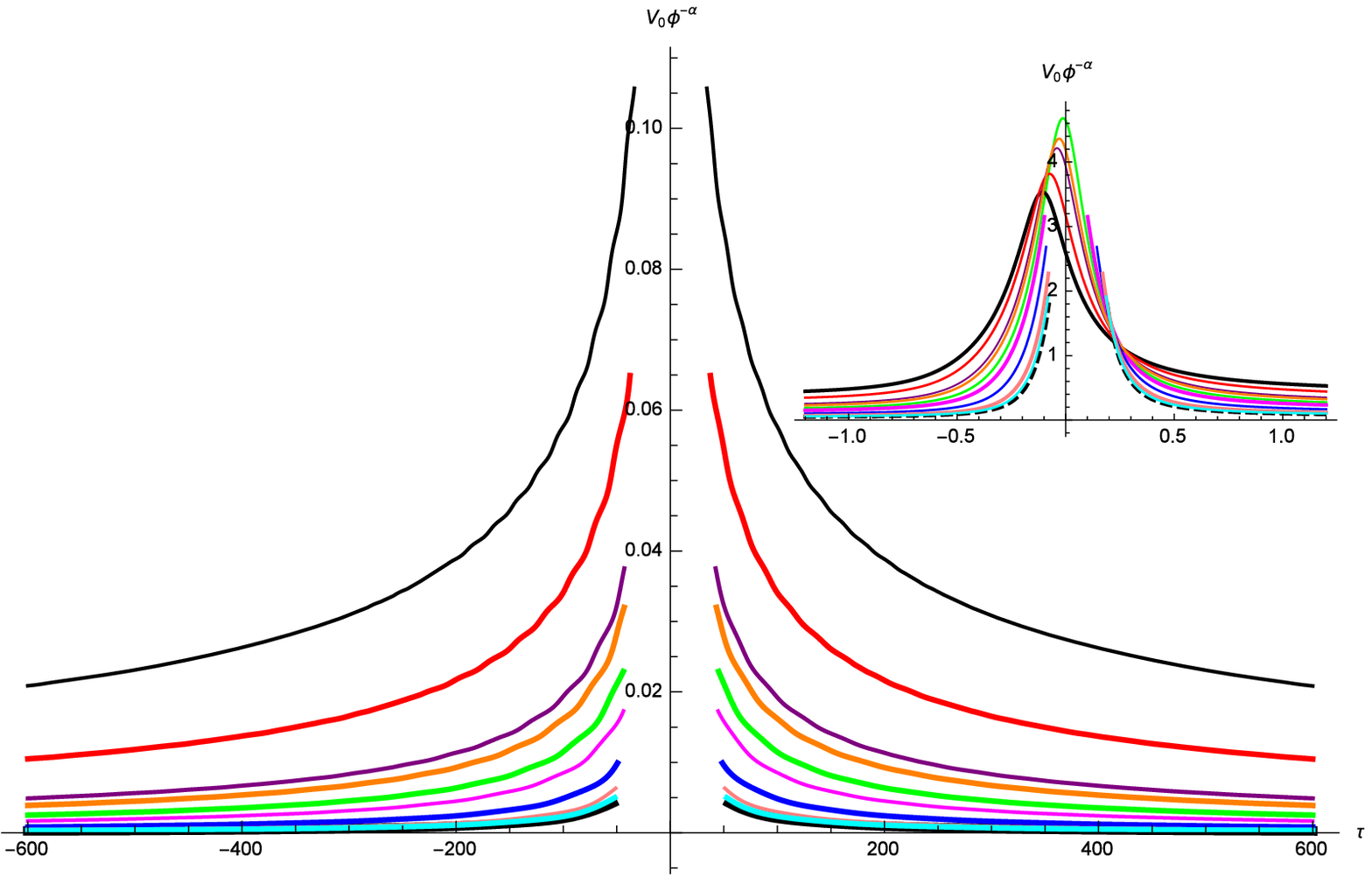}
\caption{fig.7b: $V(\phi)$ versus $\tau$}
\label{fig:sub2}
\end{subfigure}
\caption{The plot of $a(\tau)$ and $V(\phi)$ with $\tau$ are shown respectively in fig.7a and fig.7b  using initial conditions $\Big(a(0.1)=1.2, a'(0.1)=-0.2, \phi(0.1)=0.5, \phi'(0.1)=2.2\Big)$ for $V_0\phi^{-\alpha}$ potential with $\gamma=-1$, $V_0=1$, $K=1$, $\kappa=1$, $\nu=0.7$ and $\Lambda_0=1$. The  values of the index $\alpha$ in the plots are $\alpha = 2.6,~ \frac{5}{2•},~ \frac{7}{3•},~ 2,~ \frac{5•}{3•},~ \frac{3}{2•},~  \frac{4}{3•},~ \frac{5}{4•},~ 1,~ \frac{4}{5•}$ respectively in the black dotted, cyan, pink, blue, magenta, green, orange, purple, red and black curves in fig.7.}
\label{figure 10}
\end{figure}

\subsection{Appendix-II: Expression of $a(\tau)$ using fit of numerical solution with polynomial in $\tau$:}
Again the fit of numerical solution of $a(\tau)$ using polynomial of odd and even power of $\tau$ within $-50\leq \tau \leq 50$ for $V_0\phi^{-2}$ potential with initial condition of fig.3a gives
\begin{multline}\label{g2.42}
a(\tau)=2.23419 -0.071019 \tau +0.379634 \tau ^2+ 0.00413109 \tau ^3-0.00800463
\tau ^4-0.0000748033 \tau ^5+
0.000090918 \tau ^6 \\+7.5182179 *10^{-7} \tau ^7-6.3204586* 10^{-7}
\tau ^8-4.8473637*10^{-9} \tau ^9+ 2.87390304 *10^{-9} \tau ^{10}+\\
2.1081491*10^{-11} \tau ^{11}-8.9022224*10^{-12} \tau ^{12}-6.3825773*10^{-14} \tau ^{13}+ 1.9280816 *10^{-14} \tau ^{14}+1.3750221*10^{-16}
\tau ^{15}-\\
2.9541519*10^{-17} \tau ^{16}-2.1307452*10^{-19} \tau ^{17}+ 3.1810737*10^{-20} \tau ^{18}+ 2.3648932*10^{-22} \tau ^{19}-2.3169987*10^{-23}
\tau ^{20}-\\
1.824845*10^{-25} \tau ^{21}+ 1.0079847*10^{-26} \tau ^{22}+ 8.9222656*10^{-29} \tau ^{23}- 1.2695245*10^{-30} \tau ^{24}-1.8388679*10^{-32}
\tau ^{25}-\\
1.0695908*10^{-33} \tau ^{26}-6.5039467*10^{-36} \tau ^{27}+ 4.3551887*10^{-37} \tau ^{28}+  4.7491242*10^{-39} \tau ^{29}+ 9.624324*10^{-41}
\tau ^{30}-\\
3.0447219*10^{-44} \tau ^{31}-7.70942*10^{-44} \tau ^{32}-7.5475389*10^{-46} \tau ^{33}-1.049338*10^{-47} \tau ^{34}+ 1.1531204*10^{-49}
\tau ^{35}+\\
1.2831719*10^{-50} \tau ^{36}+ 1.0942404*10^{-52} \tau ^{37}+ 1.2491877*10^{-54} \tau ^{38}- 3.6785252*10^{-56} \tau ^{39}- 2.2327906*10^{-57}
\tau ^{40}-\\
1.1648022*10^{-59} \tau ^{41}+ 7.6027120*10^{-63} \tau ^{42}+ 1.0323276*10^{-62} \tau ^{43}+ 4.0566396*10^{-64} \tau ^{44}- 2.8878199*10^{-66}
\tau ^{45}-\\
1.5074807*10^{-67} \tau ^{46}+ 3.8790993*10^{-70} \tau ^{47}+ 2.3591084*10^{-71} \tau ^{48}- 2.1208468*10^{-74} \tau ^{49}- 1.4339072*10^{-75}
\tau ^{50}
\end{multline}

The fit of numerical solution of $a(\tau)$ using polynomial of even power of $\tau$  within $-50\leq \tau \leq 50$ for $V_0 \phi^{-2}$ potential with initial condition of fig.3b gives
\begin{multline}\label{g2.43}
a(\tau)=2.23756 +0.37757 \tau ^2-7.89409 *10^{-3}\tau ^4+ 8.88124 * 10^{-5} \tau^6-6.1203496 *{10}^{-7} \tau ^8+\\
2.76216675 * 10^{-9} \tau ^{10} -8.5021053* 10^{-12} \tau ^{12}
+1.8314958 * 10^{-14} \tau ^{14}-2.7929551 * 10^{-17} \tau ^{16} +\\ 2.9945401 *10^{-20}
\tau ^{18}-
2.1715504 *10^{-23} \tau ^{20} +9.3904486 *10^{-27} \tau ^{22}-1.1507882* 10^{-30} \tau ^{24}-\\1.0100342*10^{-33} \tau ^{26} +4.0420694 *10^{-37}
\tau ^{28}+
9.2378076 *10^{-41} \tau ^{30}-7.2048338 * 10^{-44} \tau ^{32}-\\1.0208739* 10^{-47} \tau ^{34}+ 1.2047031 * 10^{-50} \tau ^{36} +  1.2292825 *10^{-54}
\tau ^{38}-
2.1068493 *10^{-57} \tau ^{40} +\\3.160816 *10^{-64} \tau ^{42} +3.8512670* 10^{-64} \tau ^{44} -1.4260117* 10^{-67} \tau ^{46}\\ +2.2284504 *10^{-71}
\tau ^{48}-
1.3534363 *10^{-75} \tau ^{50}
\end{multline}
\subsection{Appendix-III: Expression of $a(\tau)$ using fit of numerical solution with polynomial in $\tau$:}
Again fit of numerical solution of $a(\tau)$ using polynomial of odd and even power of $\tau$ within $-100\leq \tau \leq 100$ for $V_0 e^{-\mu\phi}$ potential with initial condition of fig.8a gives
\begin{multline}\label{g2.44}
a(\tau)=1.53331+0.0385906 \tau +0.154504 \tau ^2 -0.000337597\tau ^3-0.000794916 \tau ^4 +1.29187*10^{-6}\tau ^5+2.53587*10^{-6}\tau ^6 \\ -2.41844*10^{-9} \tau ^7-4.7721*10^{-9}
\tau ^8+2.32258*10^{-12} \tau ^9+5.71634*10^{-12}\tau ^{10}\\
-9.25707*10^{-16}\tau ^{11}-4.60595*10^{-15}\tau ^{12}-3.20447*10^{-19}\tau ^{13}+2.58307*10^{-18}\tau ^{14}
+6.18452*10^{-22}\tau ^{15}\\
-1.02476*10^{-21}\tau ^{16}-3.77939*10^{-25}\tau ^{17}+2.868*10^{-25}\tau ^{18}+1.35796*10^{-28}\tau ^{19}
-5.48544*10^{-29}\tau ^{20}\\
-3.11725*10^{-32}\tau ^{21}+6.46997*10^{-33} \tau ^{22}+4.38532*10^{-36}\tau ^{23}-2.89697*10^{-37}\tau ^{24}
-2.78165*10^{-40}\tau ^{25}-\\
3.25138*10^{-41}\tau ^{26}-1.62832*10^{-44}\tau ^{27}+4.4815*10^{-45}\tau ^{28}+3.97427*10^{-48}\tau ^{29}
+1.26388*10^{-49}\tau ^{30}\\
-5.25962*10^{-53}\tau ^{31}-4.49476*10^{-53}\tau ^{32}-3.82583*10^{-56}\tau ^{33}-5.90006*10^{-58}\tau ^{34}
+1.8603*10^{-60}\tau ^{35}\\
+4.43797*10^{-61}\tau ^{36}+3.44701*10^{-64}\tau ^{37}+3.03464*10^{-66} \tau ^{38}-3.31041*10^{-68}\tau ^{39}
-4.59791*10^{-69}\tau ^{40}-\\
2.21357*10^{-72}\tau ^{41}+6.31847*10^{-74}\tau ^{42}+5.53121*10^{-76}\tau ^{43}+4.8687*10^{-77}\tau ^{44}
-4.03764*10^{-80}\tau ^{45}-\\
4.86454*10^{-81}\tau ^{46}+1.40241*10^{-84}\tau ^{47}+1.96883*10^{-85}\tau ^{48}-1.97833*10^{-89}\tau ^{49}
-3.05922*10^{-90}\tau ^{50}
\end{multline}

\begin{figure}
\centering
\begin{subfigure}{.45\textwidth}
\includegraphics[width=\linewidth]{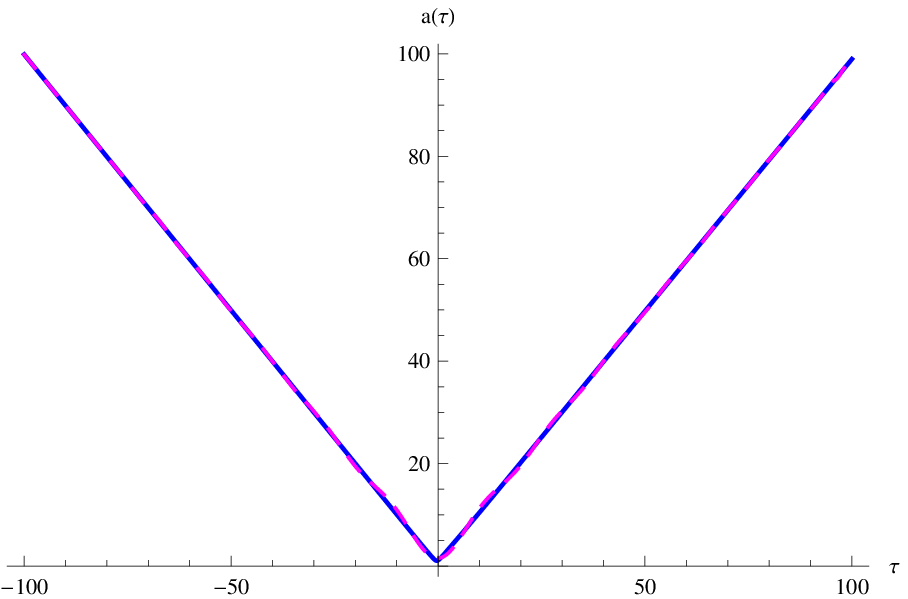}
\caption{fig.8a: Plot of $a(\tau)$ (blue curve) and the corresponding fitted dashed magenta curve using fit with polynomial of ``odd and even" power of $\tau$ for $e^{-\mu\phi}$ potential.}
\label{fig:sub1}
\end{subfigure}
\begin{subfigure}{.45\textwidth}
\includegraphics[width=\linewidth]{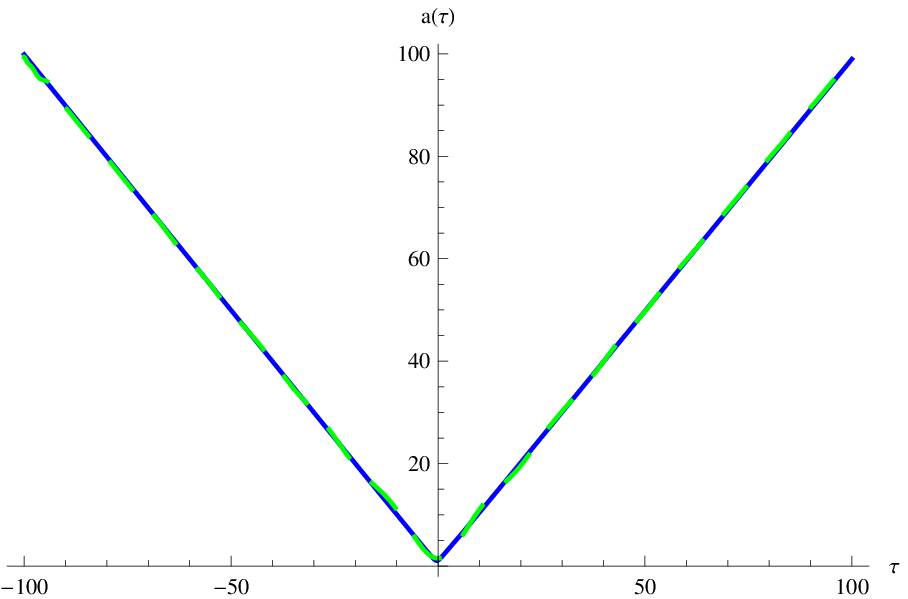}
\caption{fig.8b: The plot of $a(\tau)$ (blue curve) and the corresponding fitted dashed green curve using fit with polynomial of `` even" power of  $\tau$ for $e^{-\mu\phi}$ potential.}
\label{fig:sub2}
\end{subfigure}
\caption{The initial conditions in numerical solution of $a(\tau)$ shown in black dashed curve are $\Big(a(1)=1.9,~a'(1)=.9, ~\phi(1)=0.8,~\phi'(1)=-.4\Big)$ same as fig.1a with $\gamma=-1$, $V_0=1$, $K=1$, $\kappa=1$, $\nu=1$, $\mu=8$ and $\Lambda_0=1$.}
\label{figure 7}
\end{figure}

\begin{figure}
\centering
\begin{subfigure}{.45\textwidth}
\includegraphics[width=\linewidth]{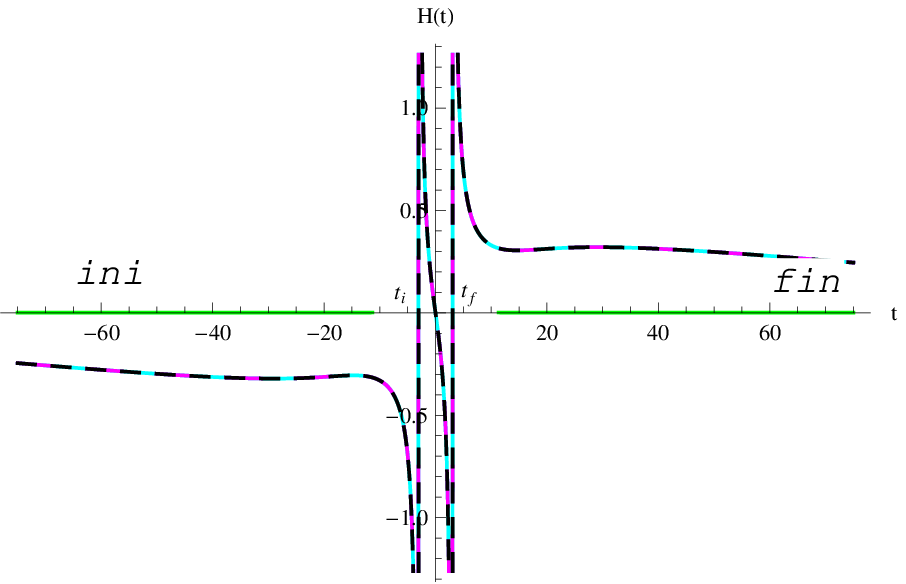}
\caption{fig.9a: Plot of $H(t)$ versus $t$ for $e^{-\mu\phi}$ potential.}
\label{fig:sub1}
\end{subfigure}
\begin{subfigure}{.45\textwidth}
\includegraphics[width=\linewidth]{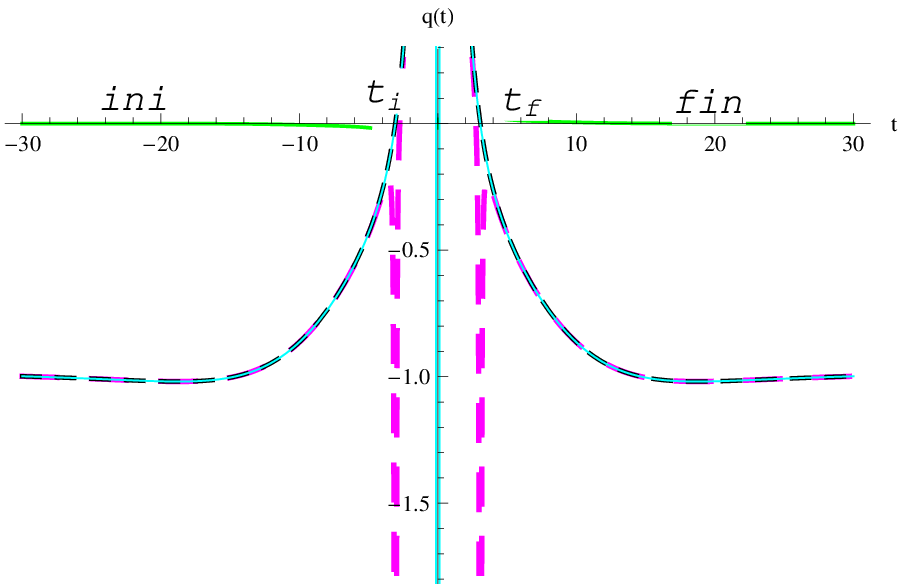}
\caption{fig.9b: Plot of $q(t))$ versus $t$ for $e^{-\mu\phi}$ potential.}
\label{fig:sub2}
\end{subfigure}
\caption{The evolution of Hubble parameter $H(t)$ and deceleration parameter $q(t)$ with $t$ are shown respectively in fig.9a and fig.9b. Real and imaginary parts of $H(t)$ and $q(t)$ obtained from the fit of $a(\tau)$ with polynomial of ``odd and even" power of $\tau$ are shown respectively in cyan and green curves. Again fit of $a(\tau)$ with polynomial of ``even" power of $\tau$ leads to real $H(t)$ and $q(t)$, which are shown by black dashed curves.The magenta dashed curves for $H(t)$ and $q(t)$ are plotted from the absolute value of a(t) using (38).}
\label{figure 7}
\end{figure}

The fit of numerical solution of $a(\tau)$ using polynomial of even power of $\tau$  within $-100\leq \tau \leq 100$ for $V_0 e^{-\mu\phi} $ potential with initial condition of fig.8b gives
\begin{multline}\label{g2.45}
a(\tau)=1.5275+0.1514\tau ^2-0.00076\tau ^4+2.3941*10^{-6}\tau^6-4.4594*10^{-9}\tau ^8\\
+5.2963*10^{-12}\tau ^{10} -4.23510* 10^{-15} \tau ^{12}
+ 2.35826* 10^{-18} \tau ^{14}- 9.29160* 10^{-22} \tau ^{16} +\\ 2.58246 *10^{-25}
\tau ^{18}-4.90201*10^{-29} \tau ^{20} +5.72424 *10^{-33} \tau ^{22}-2.49059* 10^{-37} \tau ^{24}-\\2.94192*10^{-41} \tau ^{26} +3.92669*10^{-45}
\tau ^{28}+1.19820*10^{-49}\tau ^{30}-3.958594 * 10^{-53} \tau ^{32}-\\5.99864* 10^{-58} \tau ^{34}+3.91734* 10^{-61} \tau ^{36} +3.38699 *10^{-66}
\tau ^{38}-4.06910*10^{-69} \tau ^{40} +\\5.00345 *10^{-74} \tau ^{42} +4.33577* 10^{-77} \tau ^{44} -4.29249* 10^{-81}\tau ^{46} +1.72956 *10^{-85}
\tau ^{48} -2.67925*10^{-90} \tau ^{50}
\end{multline}

\pagebreak

\end{document}